\documentclass[10pt,twocolumn,english,aps,prl,nofootinbib,showkeys,preprintnumbers]{revtex4}
\usepackage[utf8]{inputenc}
\usepackage{graphicx}
\usepackage{amsmath}
\usepackage{amssymb}
\usepackage{amsfonts}
\usepackage{lscape}
\usepackage{hyperref}
\usepackage{url}
\usepackage{epsfig}
\usepackage{color}
\usepackage{bm}
\usepackage{tabularx}

\def\ga{\mathrel{\raise.3ex\hbox{$>$\kern-.75em\lower1ex\hbox{$\sim$}}}}
\def\la{\mathrel{\raise.3ex\hbox{$<$\kern-.75em\lower1ex\hbox{$\sim$}}}}

\newcommand{\be}{\begin{equation}}
\newcommand{\ee}{\end{equation}}
\newcommand{\bea}{\begin{eqnarray}}
\newcommand{\eea}{\end{eqnarray}}
\newcommand{\nn}{\nonumber}

\def\Msun{M_\odot}
\def\Mchirp{{\cal M}_c}

\def\L{{\cal L}}
\def\P{{\cal P}}
\def\chieff{\chi_{\rm eff}}
\def\af{a_{\rm f}}

\long\def\commentout#1{}

\begin{document}

\title{Bayesian analysis of the spin distribution of LIGO/Virgo black holes}

\author{Juan Garc\'ia-Bellido$^a$}
\email{juan.garciabellido@uam.es}

\author{Jos\'e Francisco Nu\~no Siles$^a$}
\email{josef.nunno@estudiante.uam.es}

\author{Ester Ruiz Morales$^b$}
\email{ester.ruiz.morales@upm.es\\
ORCID:\,0000-0002-0995-595X\\}

\affiliation{
$^a$Instituto de F\'isica Te\'orica UAM-CSIC, Universidad Auton\'oma de Madrid,
Cantoblanco, 28049 Madrid, Spain\\
$^b$Departamento de F\'isica, ETSIDI, Universidad Polit\'ecnica de Madrid, 28012  Madrid, Spain
}

\date{\today}

\preprint{IFT--UAM/CSIC--18--76, CERN--TH--2018--158}

\begin{abstract}
Gravitational wave detection from binary black hole (BBH) inspirals has become routine thanks to the LIGO/Virgo interferometers. The nature of these black holes remains uncertain. We study here the spin distributions of LIGO/Virgo black holes from the first catalogue GWTC-1 and the first four published BBH events from run O3. We compute the Bayes evidence for several independent priors: flat, isotropic, spin-aligned and anti-aligned. We find strong evidence for low spins in all of the cases, and significant evidence for small isotropic spins versus any other distribution. When considered as a homogeneous population of black holes, these results give support to the idea that LIGO/Virgo black holes are primordial. 
\end{abstract}


\maketitle



\section{Introduction}

The regular detection with laser interferometers~\cite{LIGOScientific:2018mvr,Abbott:2020uma,LIGOScientific:2020stg,Abbott:2020khf,Abbott:2020tfl} of gravitational wave (GW) events from BBH mergers has opened a new window into the universe, and in particular to the exploration of the nature of black holes.

Before the first BBH detections by LIGO, stellar black holes with masses in the range $5-15~\Msun$ had been detected as components of X-ray binaries, only a few intermediate mass black holes (IMBH) were known with masses above $100\,\Msun$, while supermassive black holes (SMBH) were known to exist at the centers of all galaxies. The origin of black holes in such a large range of masses remains a mystery, and one fascinating possibility is that part of these black holes are primordial in origin~\cite{Clesse:2015wea,Garcia-Bellido:2017fdg}. In fact, soon after the first detection of black hole mergers by LIGO~\cite{Abbott:2016blz}, there were claims of their primordial nature~\cite{Bird:2016dcv,Clesse:2016vqa,Sasaki:2016jop}. Since then, the best scenario consistent with all observational constraints so far~\cite{Carr:2019kxo} is that of spatially clustered and broad-mass distributed primordial black holes~\cite{Clesse:2017bsw,Garcia-Bellido:2017fdg}.

As the number of binary black hole merger events detected with GW interferometers increases, a new population of black holes is emerging with unexpected properties in terms of their masses and spins. These properties significantly differ from previous black holes detected through X-rays, via stellar dynamics around SMBH at the center of galaxies, or via gravitational lensing effects for IMBH. When the range of masses and distances accessible by the GW interferometers improved, events like GW190425 and GW190521 appeared with BH masses in the lower and upper mass gaps, challenging existing astrophysical BH formation models. Moreover, events with small mass ratios $q\ll1$, like GW190814, are also difficult to generate in stellar binary formation models, due to the expected mass transfer among the binary components, see however~\cite{Zevin:2020gma}.

One of the most striking features of this new population of black holes detected in GW events is that they all seem to have a small spin. Although the individual spin of each of the black holes in the binary is poorly constrained,\footnote{It is worth pointing out, however, that the best determined spin for a single black hole is that of the  massive companion of GW190814, with $S_1<0.07$ at 90\% c.l., thanks to its particularly low mass ratio $m_2/m_1=0.11$.} a derived quantity called the {\em effective} spin, $\chieff$, can be well inferred from the GW waveform. The O1+O2 events \cite{LIGOScientific:2018mvr} plus the four run O3 published events~\cite{Abbott:2020uma,LIGOScientific:2020stg,Abbott:2020khf,Abbott:2020tfl} show that, in almost all the BBHs mergers events, the inferred $\chieff$ posteriors cluster around zero and are narrowly peaked. This single observation already constrains some BBH formation models, because even if it is possible to generate a single merger event with very small $\chieff$ in almost all physical scenarios of BBH formation, the fact that {\em the whole population} of BBH mergers have very small $\chieff$ cannot be explained by models involving high aligned spins of the two BH's in the binary.

Even before the first BBH merger detections, the effective spin had already been considered an optimal variable to discriminate among formation channels. By simulating populations of BBH mergers with different intrinsic spin values and spin alignments, the authors in~\cite{Vitale:2015tea,Talbot:2017yur} showed that the effective spin could be relevant to distinguish the astrophysical environment in which these binary systems formed, since binaries evolved by dynamical interactions were expected to have their spins isotropically oriented, while those coming from the evolution of an isolated binary star system were more likely to have their spins highly aligned (see however~\cite{Belczynski:2017gds} for updates on these models.).

The magnitude of the intrinsic spin of each black hole in the binary, though much poorly constrained from the data than the effective spin, is important to determine how these black holes were formed, and in particular to distinguish a possible stellar versus primordial origin. Stellar black holes are expected to have nonzero spin, due to conservation of angular momentum. On the contrary, since the size of their Schwarzschild radius is essentially identical to the size of the causal horizon at the moment of their formation, primordial black holes are all expected to have zero or very near zero spin at formation~\cite{DeLuca:2019buf}, although subsequent accretion could enhance it slightly~\cite{DeLuca:2020bjf}.

On the other hand, it has been argued that the more massive LIGO/Virgo BH could originate from previous mergers~\cite{Gerosa:2017kvu,Fishbach:2017dwv}. However, it is known that the spin distribution of second generation black holes is peaked very far from zero, near $S\sim0.686$~\cite{Barausse:2009uz}, and this is in disagreement with the fact that most of the massive BH in LIGO/Virgo seem to have very low spin. In a careful analysis, Ref~\cite{Kimball:2020opk} showed that GWTC-1 BBH catalog is consistent with having no hierarchical mergers.

When the first detections of the O1 run became available, Farr et al.~\cite{Farr:2017uvj,Farr:2017gtv} analysed the discriminating power of the $\chieff$ distribution with the first four GW LIGO/Virgo events, assuming equal masses ($q=m_2/m_1=1$) for all events and approximating the posterior $\chieff$ distributions by Gaussians. They compared the odds ratios for several models of spin (modulus and orientations) of the underlying BH population, finding a preference for either a population with an isotropic spin distribution or with low intrinsic spin values of the merging black holes. A more realistic analysis of the first six LIGO/Virgo mergers was made in~\cite{Tiwari:2018qch}, including Bayesian methods and taking into account the $q-\chieff$ correlations, reaching the general conclusion that highly spinning black holes were disfavoured against low spins, see also~\cite{Fernandez:2019kyb}. 

With the publication of the GWTC-1 catalog, the LIGO/Virgo collaboration (LVC) made a population analysis of the mass, redshift and spin properties of the ten BBH mergers detected in O1+O2 runs~\cite{LIGOScientific:2018jsj}.\footnote{A hierarchical Bayesian analysis of spin distributions was also performed in~\cite{Miller:2020zox}. Another Bayesian analysis, studying mass distributions and merger rates but not spins, was done in~\cite{Hall:2020daa}.} Under different assumptions for the parameters of the population models, they observed a common trend of the inferred distribution for the BH spins to decrease with increasing spin magnitude, but were not able to place strong constraints on spin orientations, concluding that black hole spin measurements were not informative enough at that moment to discern between isotropic and aligned orientation distribution via $\chieff$.

In this work, we improve such discriminating power by incorporating the projection along the total angular momentum of the spin of the final BH formed after the merger $\af$ as another variable of the BBH merger population analysis. The effective spin and the final spin can be measured independently, since $\chieff$ is inferred from the inspiral part of the waveform while $\af$ can be determined also from the ringdown part. We follow a multivariate approach keeping all correlations among the variables ($q,\,\chieff,\,\af$) of each event, and show that this significantly improves the Bayesian evidence for an underlying population of black hole components with small spin magnitude and isotropic orientation.

There are still many uncertainties on the full spin (magnitude and orientation) distribution of the different BBH formation channels, since the computational models used to predict these properties depend on many assumptions about poorly understood environmental conditions related to the formation and evolution of the binaries. There are also many unknown aspects related to the mass, spatial distribution and dynamics of BBH of primordial origin. For this reason, instead of considering an specific astrophysical or primordial model for the mass and spin distributions, we have chosen general hypothesis on the underlying distributions of black hole spin magnitudes and orientations and compare the different hypothesis in a full Bayesian analysis.

In Section 1, we describe the spin variables and parameters that will be considered in our analysis of the BBH population. In Section 2, we describe the method employed in calculating the Bayesian evidence from the published parameter estimation samples, and define the population hypotheses together with their priors for the parameters considered in our analysis. In Section 3 we use Bayesian methods to obtain the evidence for each hypothesis and the corresponding Bayes ratios. Section 4 is devoted to the hierarchical modelling method that allows us to infer a posterior distribution for the spin magnitude of the underlying BH population. In Section 5 we give our conclusions.

\section{Spin observables from BBH events}

In this section we will describe the observables relevant to the analysis of spins in LVC black hole binaries. The main observables that we will consider are the weighted-averaged effective projected spin, $\chieff$, and the the projection along the total angular momentum of the final spin $\af$. We could have also chosen the effective precession spin parameter, $\chi_p$, but this quantity is much worse measured in the LVC events published so far and is mainly determined by the prior. 

\subsection{The effective spin}

A derived physical quantity that can be measured very well by LIGO from the waveform templates is the mass-averaged (so called effective) projected spin,
\be\label{eq:chieff}
\chieff = \frac{m_1\,\chi_1 + m_2\,\chi_2}{m_1+m_2} =
\frac{\chi_1 + q\,\chi_2}{1+q} \,, \vspace{1mm}
\ee
where $\chi_i\equiv(\vec S_i\cdot\vec L)/(|\vec L|\,m_i^2)$ is the individual mass-weighted projection of each black hole spin onto the orbital angular momentum $\vec L$ of the binary. Thus, $\chieff$ gives some information about the spin orientations of the inspiralling black holes w.r.t. the orbital angular momentum. Here $q=m_2/m_1$ is the binary mass ratio, in the range $q\in[0,1]$.

A derived quantity that will be useful is the so-called ``chirp mass",
\be\label{eq:chirpmass}
\Mchirp = \frac{(m_1\,m_2)^{3/5}}{(m_1+m_2)^{1/5}} = 
m_1\,\frac{q^{3/5}}{(1+q)^{1/5}}\,,
\ee
which gives the mass ratio
\be\label{eq:q}
q = \frac{3\alpha\,(2/3)^{2/3}+(9\alpha+\sqrt{81\alpha^2-12\alpha^3})^{2/3}}
{3\,(2/3)^{1/3}(9\alpha+\sqrt{81\alpha^2-12\alpha^3})^{1/3}}\,,
\ee
where $\alpha=(\Mchirp/m_1)^5$.

\begin{figure*}
\centering 
\includegraphics[width=0.32\textwidth,angle=0]{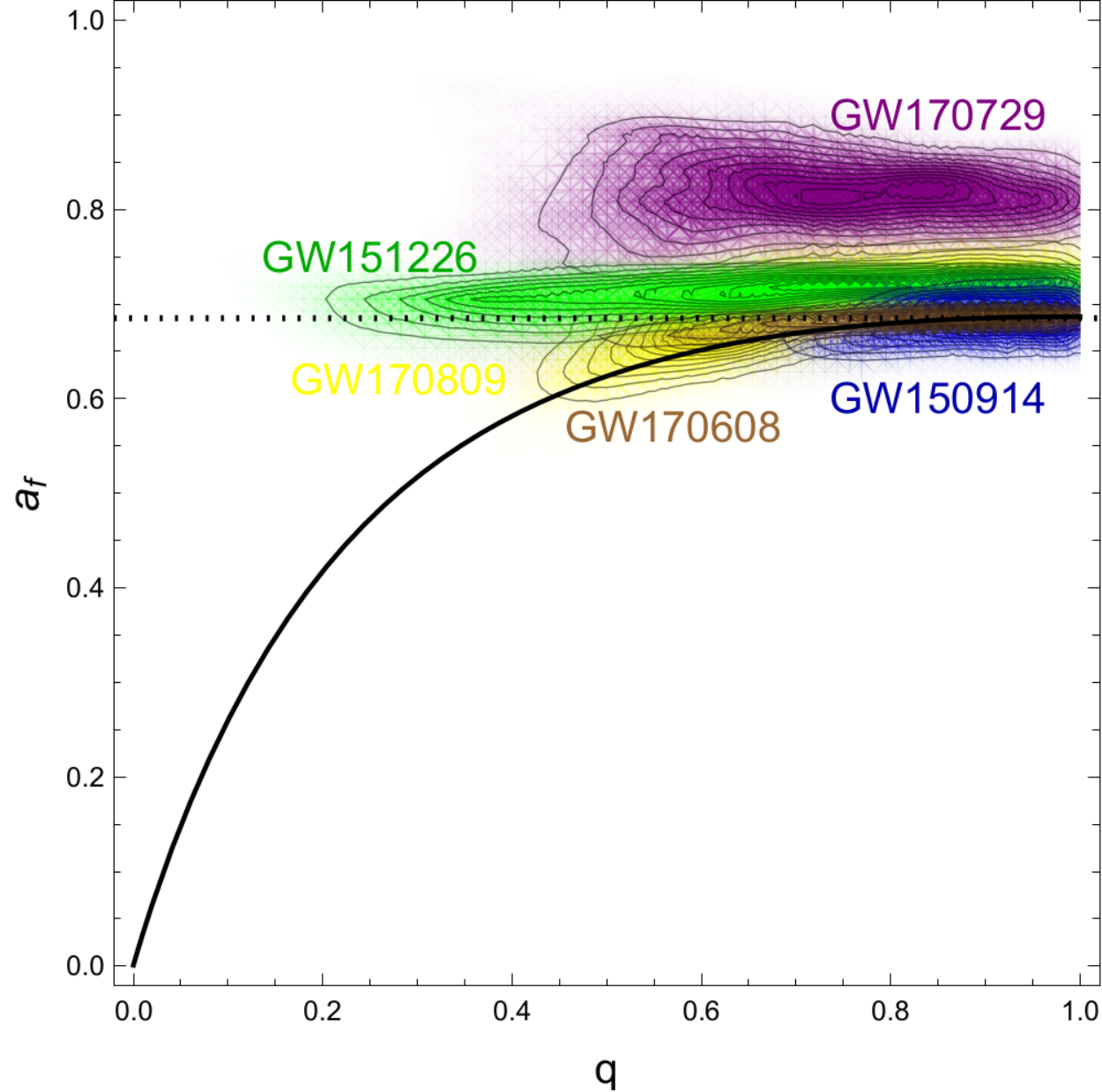}\hspace{2mm}
\includegraphics[width=0.32\textwidth,angle=0]{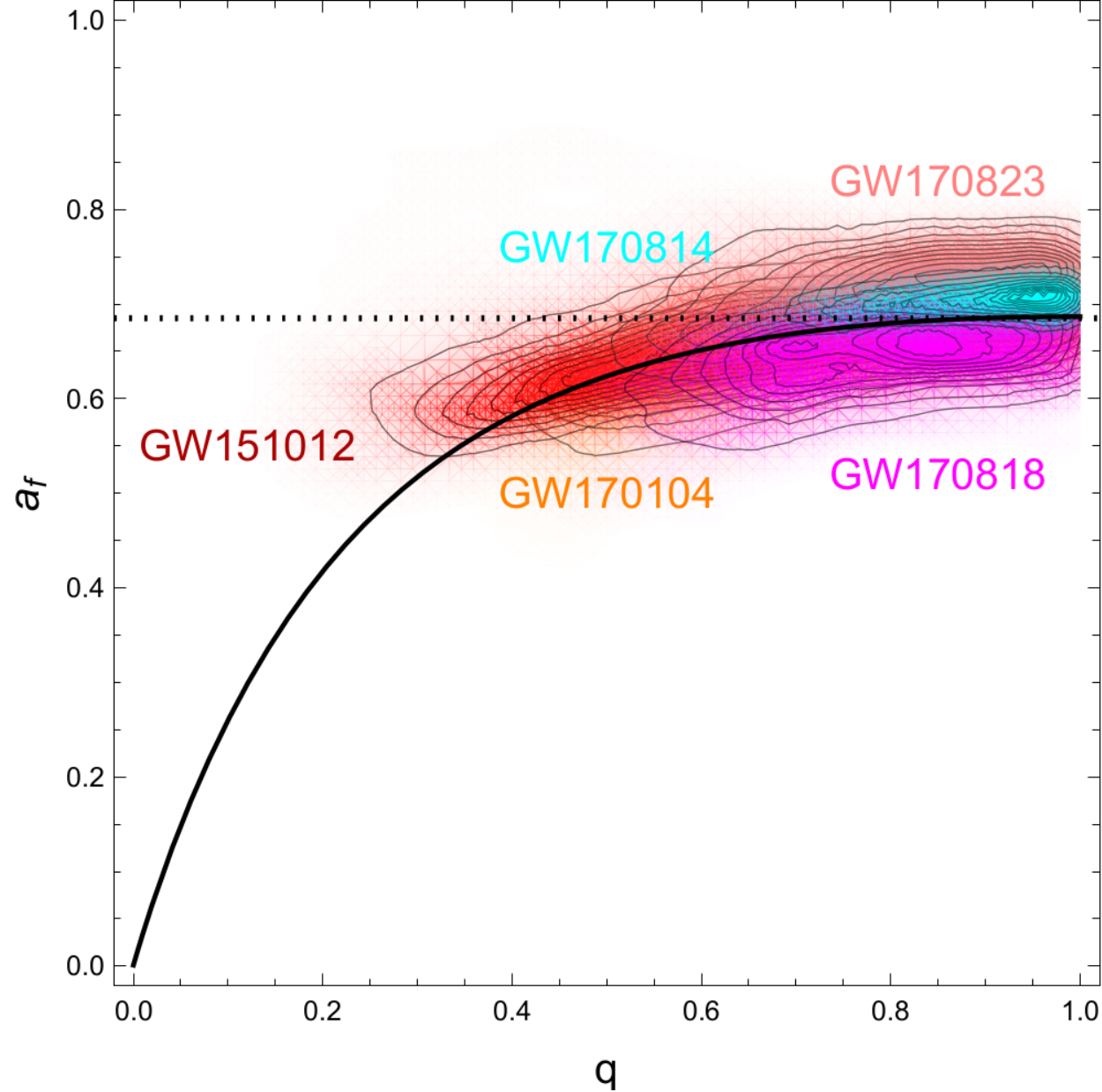}\hspace{2mm}
\includegraphics[width=0.32\textwidth,angle=0]{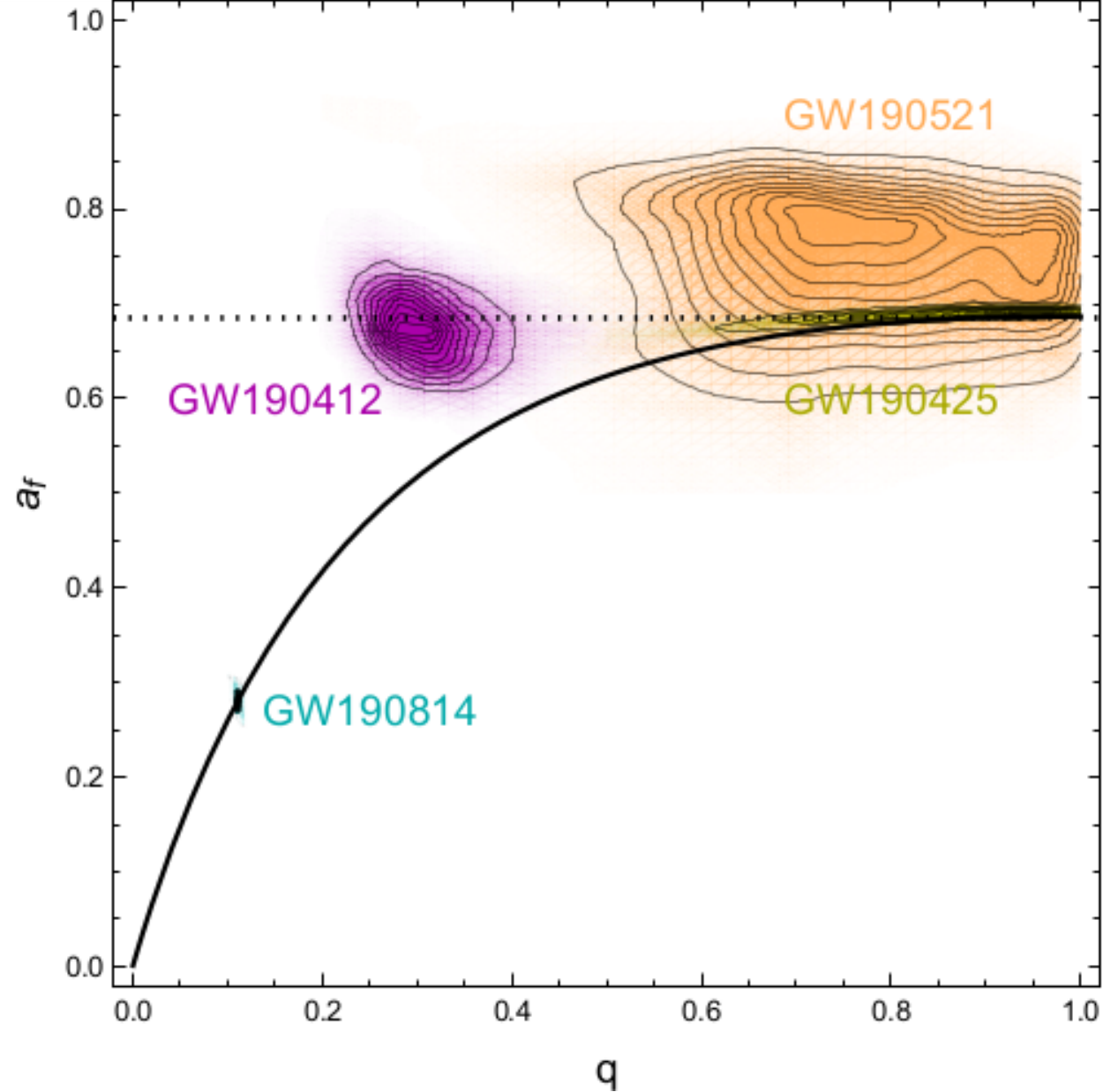}\\[2mm]
\includegraphics[width=0.32\textwidth,angle=0]{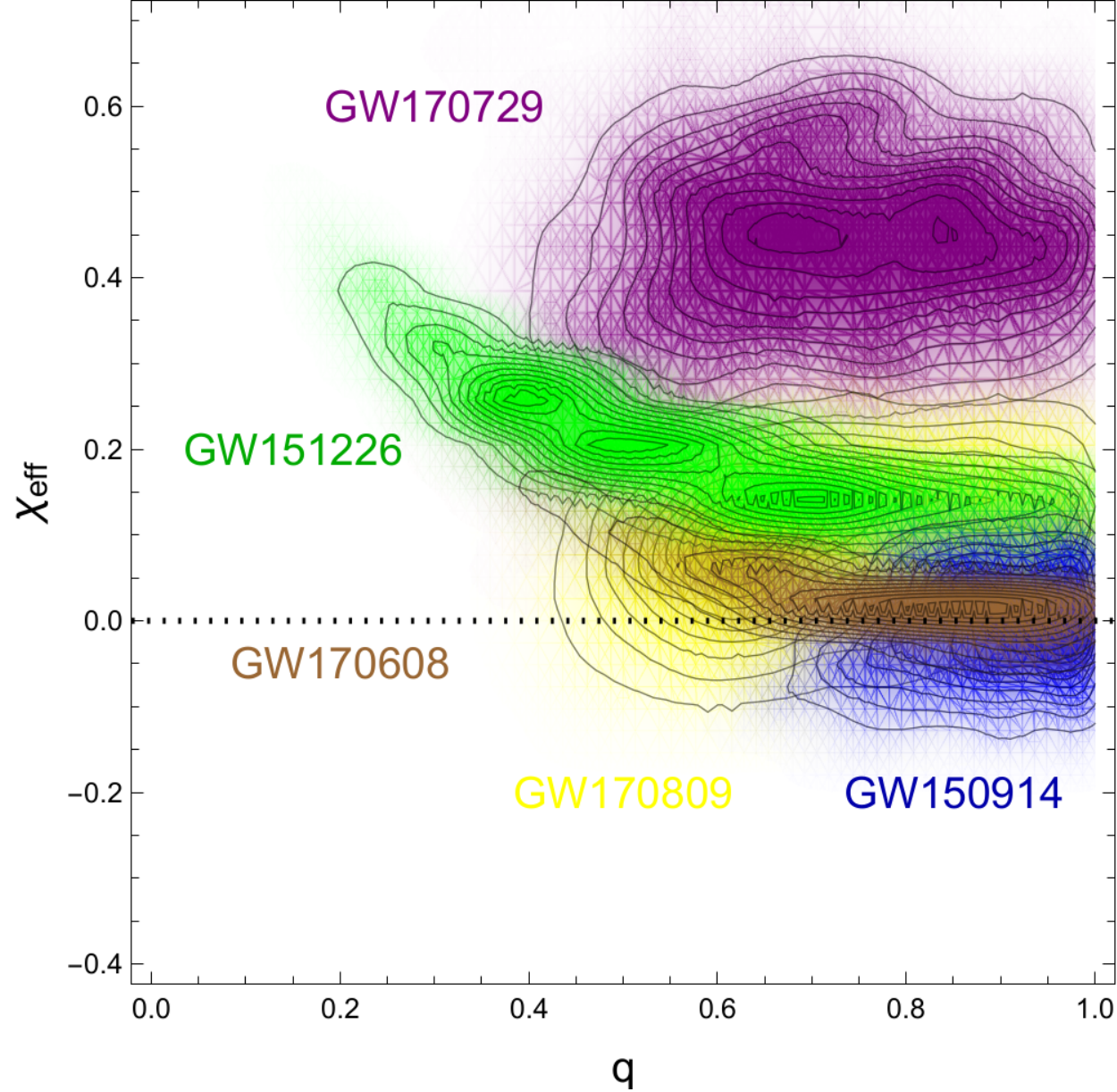}\hspace{2mm}
\includegraphics[width=0.32\textwidth,angle=0]{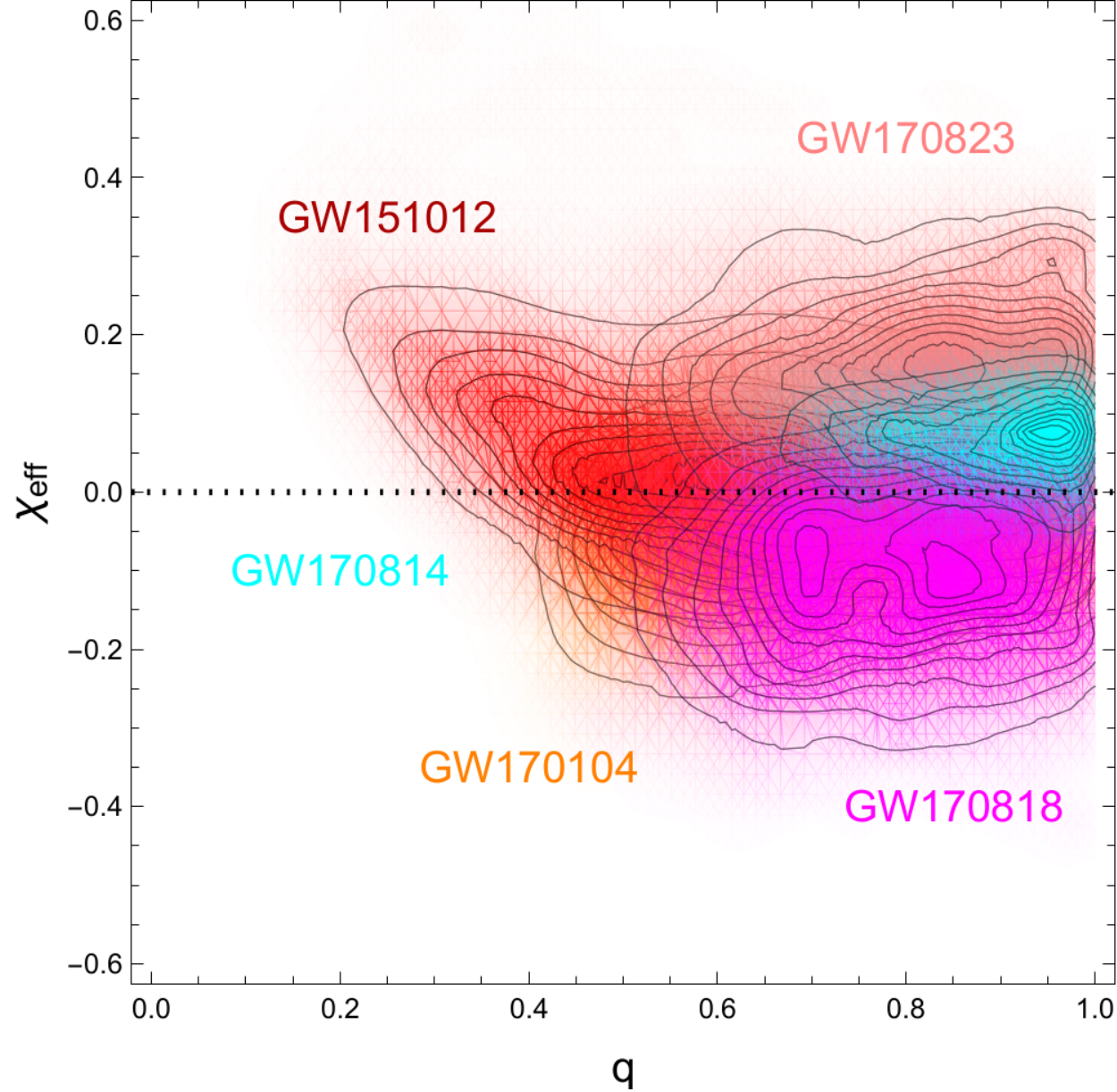}\hspace{2mm}
\includegraphics[width=0.32\textwidth,angle=0]{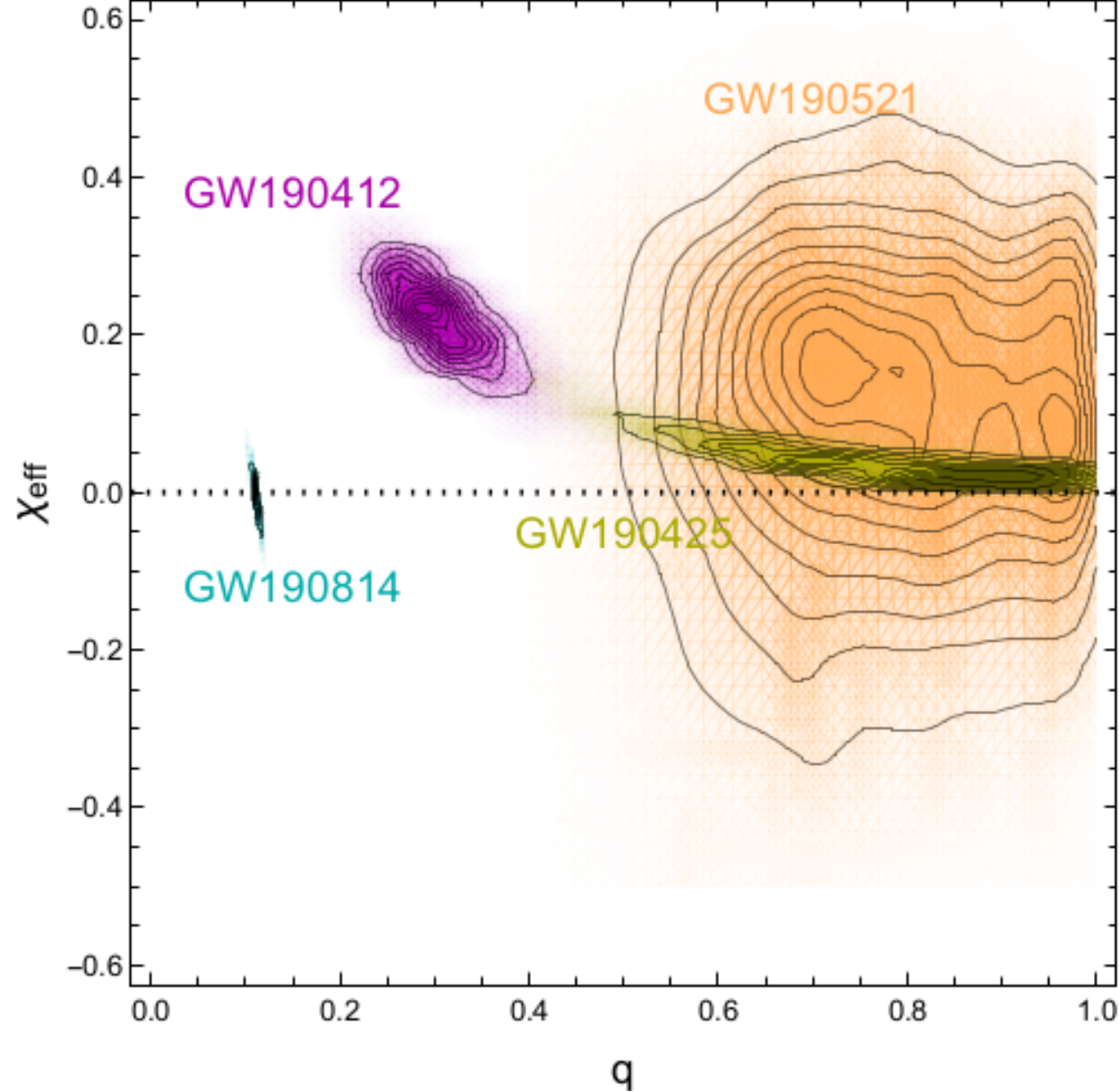}\\[2mm]
\includegraphics[width=0.32\textwidth,angle=0]{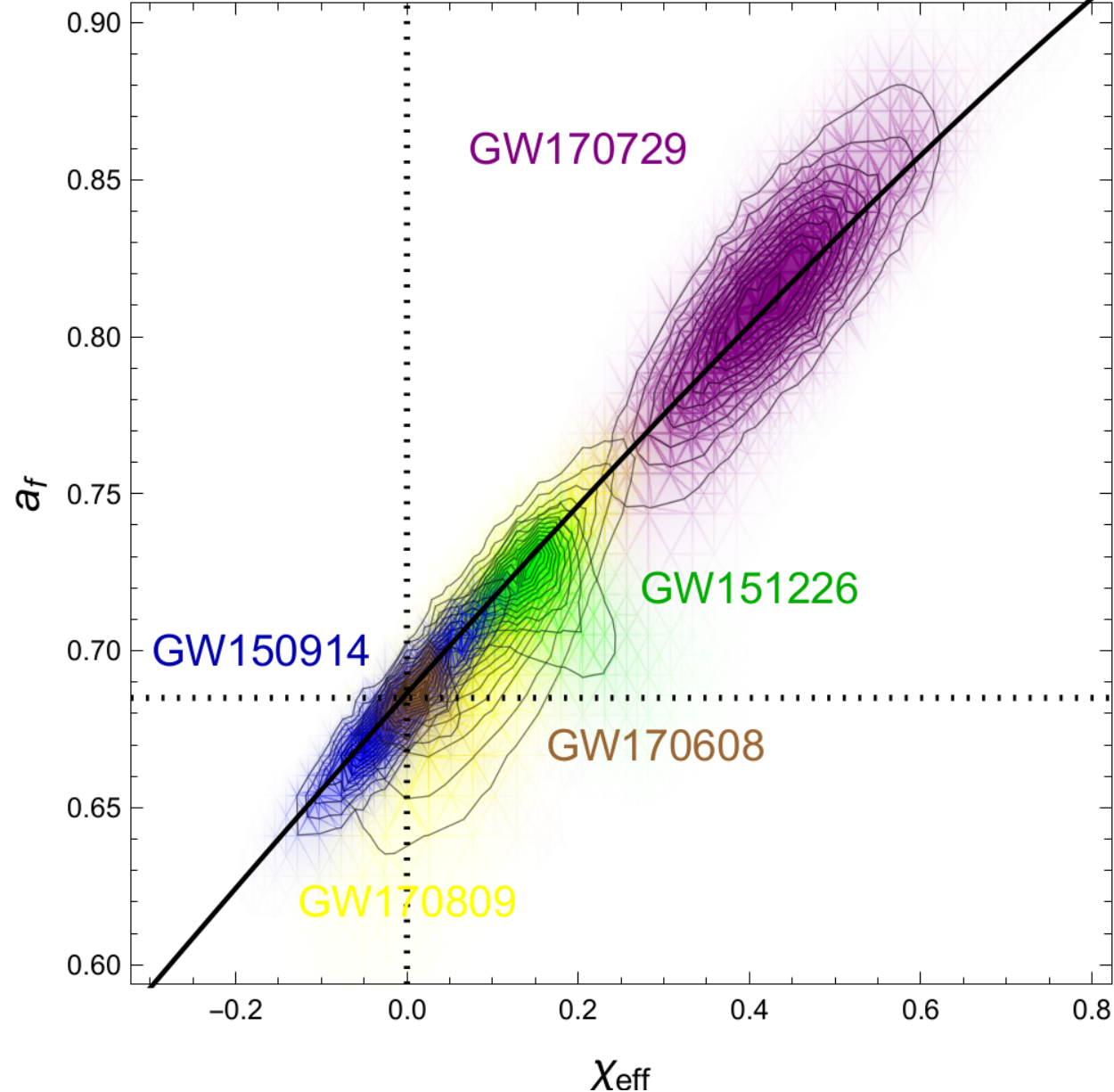}\hspace{2mm}
\includegraphics[width=0.32\textwidth,angle=0]{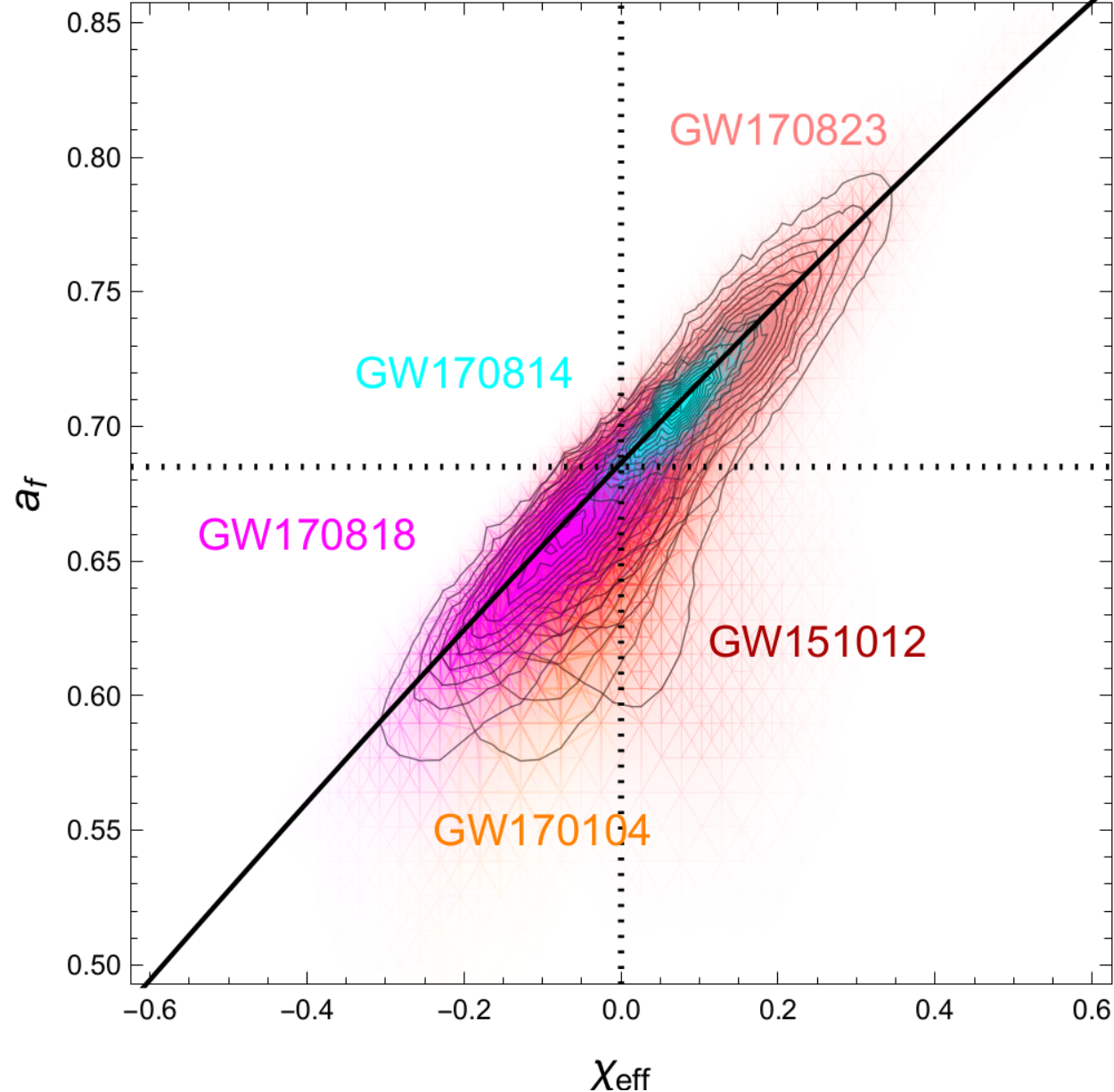}\hspace{2mm}
\includegraphics[width=0.32\textwidth,angle=0]{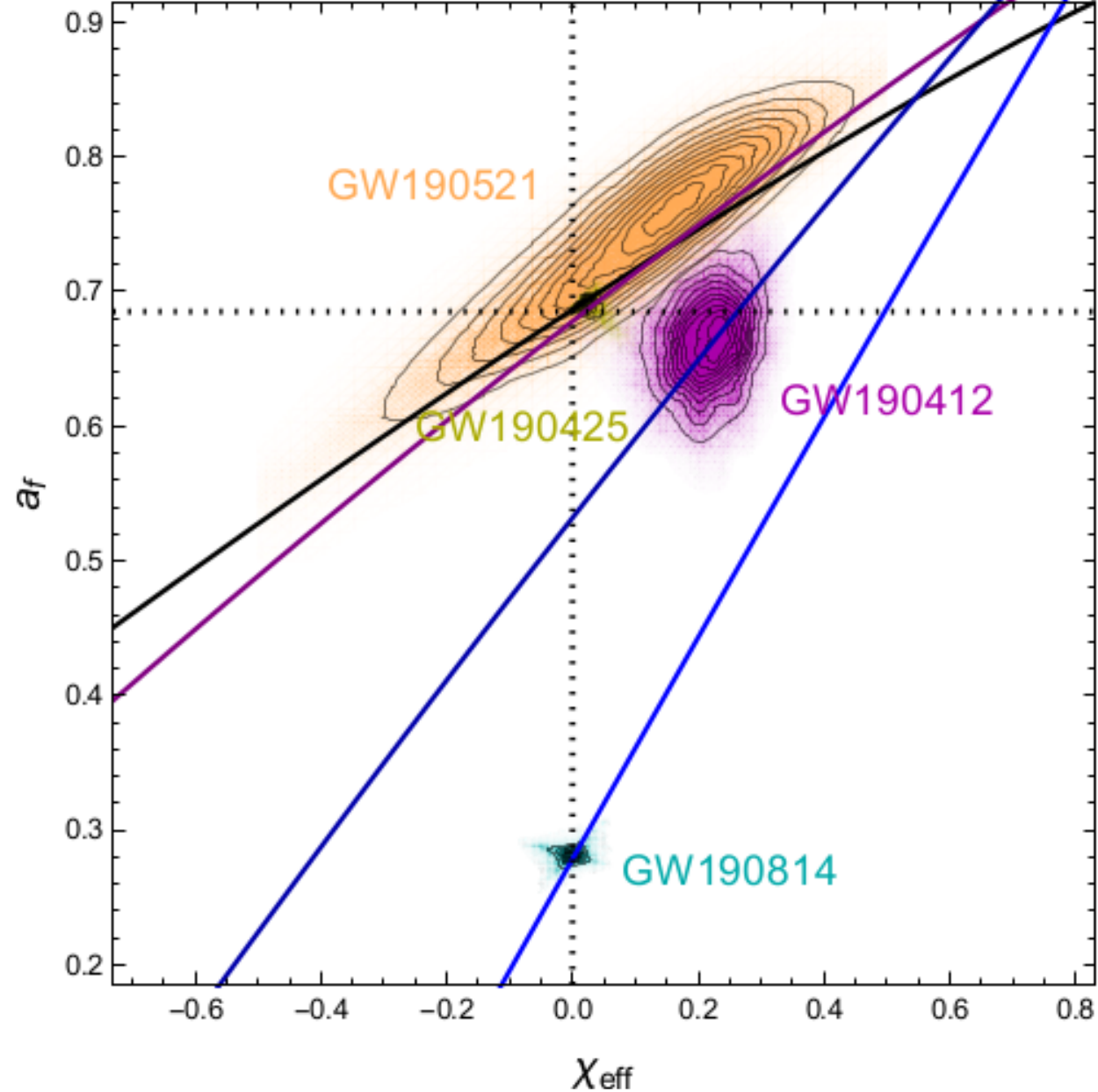}
\caption{The multidimensional LVC likelihoods projected on the planes $(q,\,\af)$  (top row), $(q,\,\chieff)$ (middle row) 
and $(\chieff,\,\af)$ (bottom row) for the ten GWTC-1 events of LIGO/Virgo (left and middle column) plus the four runO3 events (right column). The dotted lines correspond to the fixed points $\chieff=0$ and $\af=0.686$. The curves on the $(\chieff,\,\af)$ plane correspond to $q=1$ (black), $q=0.75$ (purple), $q=0.4$ (dark blue) and $q=0.11$ (light blue), respectively.
}
\label{fig:Likelihood}
\end{figure*}

\subsection{The final spin}

The second best-measured spin-related quantity in LIGO/Virgo binaries is the projection along the total angular momentum of the spin of the final black hole after merging, $\af$. 
We use the approximate expressions given in~\cite{Hofmann:2016yih} for $\af$ in the particular case of spinning but non-precessing black hole binaries. We first define some quantities and then we assemble everything together.

First we need to define $a_{\rm tot}$ and $a_{\rm eff}$ in terms of the {\em projected} spins, 
$a_i\equiv(\vec S_i\cdot\vec L)/(|\vec L|\,m_i)$,
\bea
\hspace{-2mm}a_{\rm tot} & = &  \frac{a_1 + q^2\,a_2}{(1+q)^2}  \,, 
\hspace{5mm} a_{\rm eff} = a_{\rm tot} + \xi\,\nu\,(a_1 + a_2)\,, \\[2mm] \nonumber
{\rm with} && \hspace{3mm} \nu(q) = \frac{q}{(1+q)^2}\,, \hspace{1cm} \xi=0.474046\,.
\eea
We then have to define the energy, angular momentum and size of the Innermost Stable Circular Orbit (ISCO),
\bea
&&\hspace{-2mm}E_{\rm ISCO}(a) = \sqrt{1-2/(3\,r_{\rm ISCO}(a))}\,,\\[1mm]
&&\hspace{-2mm}L_{\rm ISCO}(a) = \frac{2}{3\sqrt3}\Big(1+2\sqrt{3\,r_{\rm ISCO}(a)-2}\Big)\,,\\
&&\hspace{-2mm}r_{\rm ISCO}(a) = 3 + Z_2 - {\rm sign}(a)\sqrt{(3-Z_1)(3+Z_1+2Z_2)}\,, \nn \\[2mm]
&&\hspace{-2mm}Z_1(a) = 1+(1-a^2)^{1/3}\left((1+a)^{1/3}+(1-a)^{1/3}\right)\,,\\
&&\hspace{-2mm}Z_2(a) = \sqrt{3a^2+Z_1^2}\,.
\eea
Then the final spin is given by (the coefficients $k_{ij}$ can be found in Table~\ref{tab:LIGO})
\bea
\label{eq:af}
\af &=& a_{\rm tot} + \nu\Big(L_{\rm ISCO}(a_{\rm eff}) - 2a_{\rm tot}
(E_{\rm ISCO}(a_{\rm eff})-1)\Big)  \nn \\
&+& \nu^2 \left(k_{00} + k_{01}\,a_{\rm eff} + k_{02}\,a_{\rm eff}^2 + 
k_{03}\,a_{\rm eff}^3 + k_{04}\,a_{\rm eff}^4 \right) \nonumber \\
&+& \nu^3 \left(k_{10} + k_{11}\,a_{\rm eff} + k_{12}\,a_{\rm eff}^2 + k_{13}\,a_{\rm eff}^3 + k_{14}\,a_{\rm eff}^4 \right) \nonumber \\
&+& \nu^4 \left(k_{20} + k_{21}\,a_{\rm eff} + k_{22}\,a_{\rm eff}^2 + k_{23}\,a_{\rm eff}^3 + k_{24}\,a_{\rm eff}^4 \right) \nonumber \\
&+& \nu^5 \left(k_{30} + k_{31}\,a_{\rm eff} + k_{32}\,a_{\rm eff}^2 + k_{33}\,a_{\rm eff}^3 + k_{34}\,a_{\rm eff}^4 \right) \nn \,.
\eea

\begin{table}
\centering\renewcommand\arraystretch{1.2}
\begin{tabular}{|c|c|c|c|c|}
\hline
$-5.97723$ & $3.39221$ & $4.48865$ & $-5.77101$ & $-13.0459$ \\
\hline
$35.1278$ & $-72.9336$ & $-86.0036$ & $93.7371$ & $200.975$ \\
\hline
$-146.822$ & $387.184$ & $447.009$ & $-467.383$ & $-884.339$ \\
\hline
$223.911$ & $-648.502$ & $-697.177$ & $753.738$ & $1166.89$ \\
\hline
\end{tabular}
\caption{\label{tab:LIGO} 
The coefficients $k_{ij}$ for $i\in[0,3]$ and $j\in[0,4]$.}
\end{table}

\begin{figure*}
\centering 
\includegraphics[width=0.32\textwidth,angle=0]{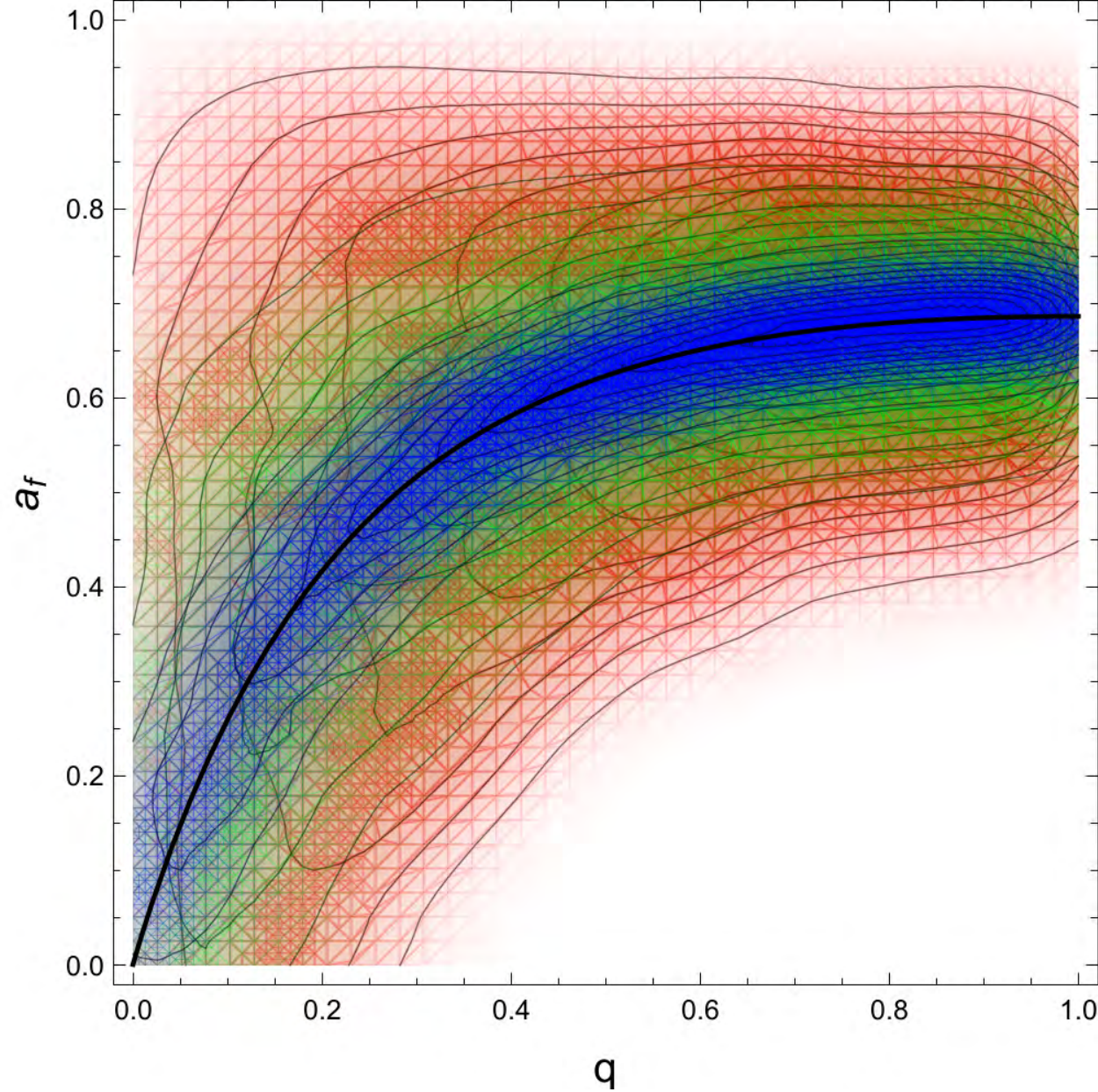}\hspace{1mm}
\includegraphics[width=0.32\textwidth,angle=0]{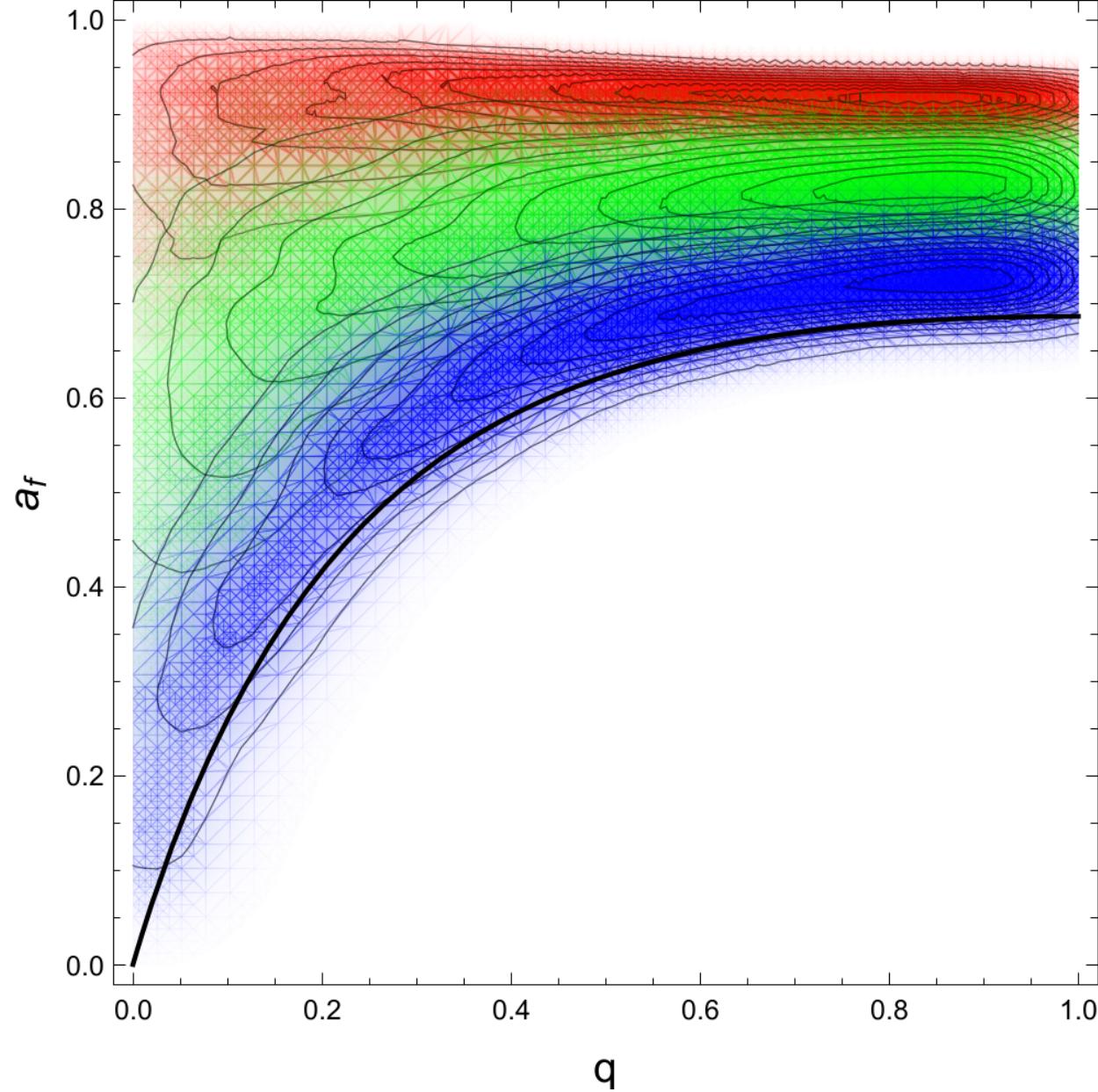}\hspace{1mm}
\includegraphics[width=0.32\textwidth,angle=0]{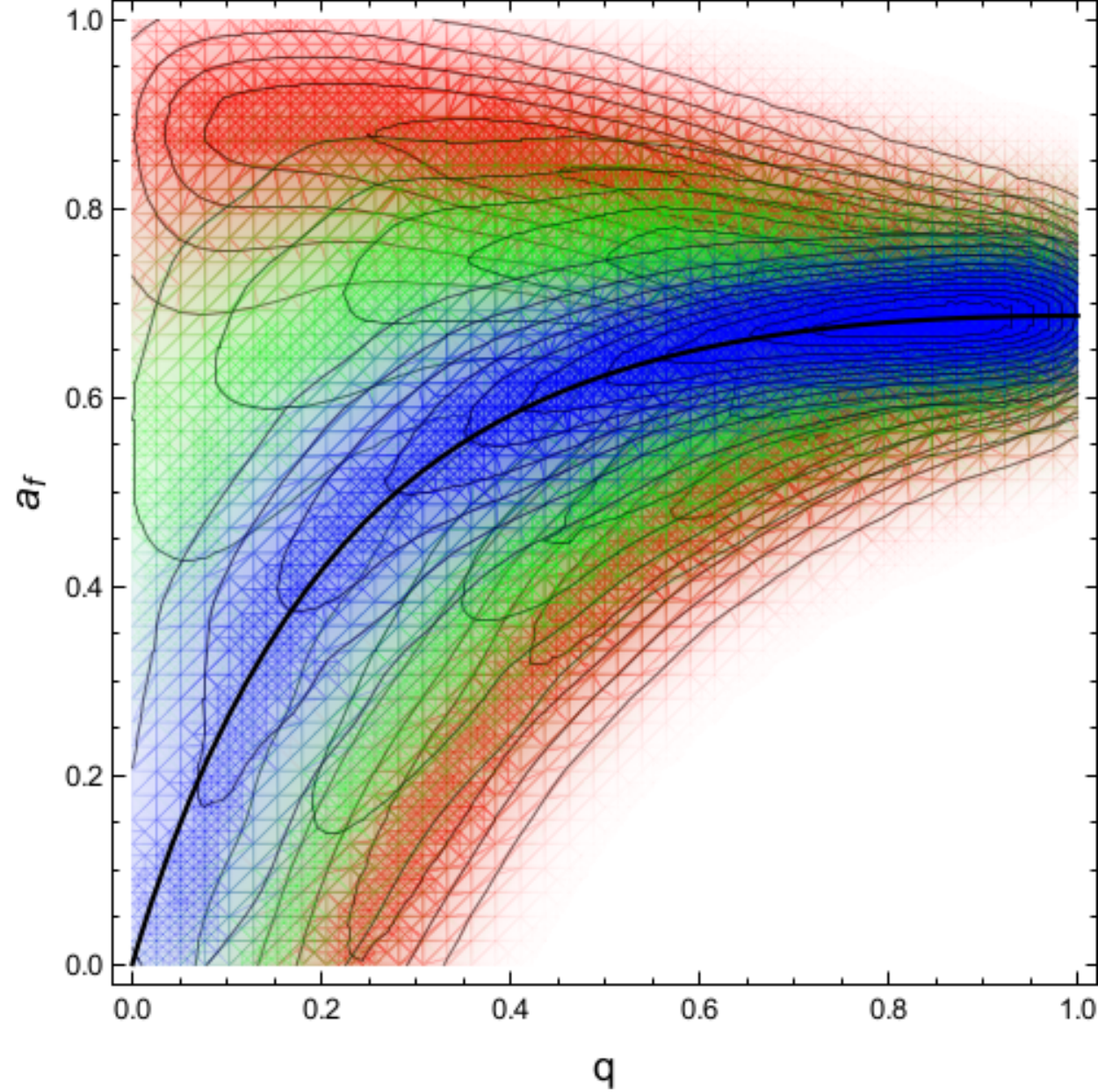}\\[2mm]
\includegraphics[width=0.32\textwidth,angle=0]{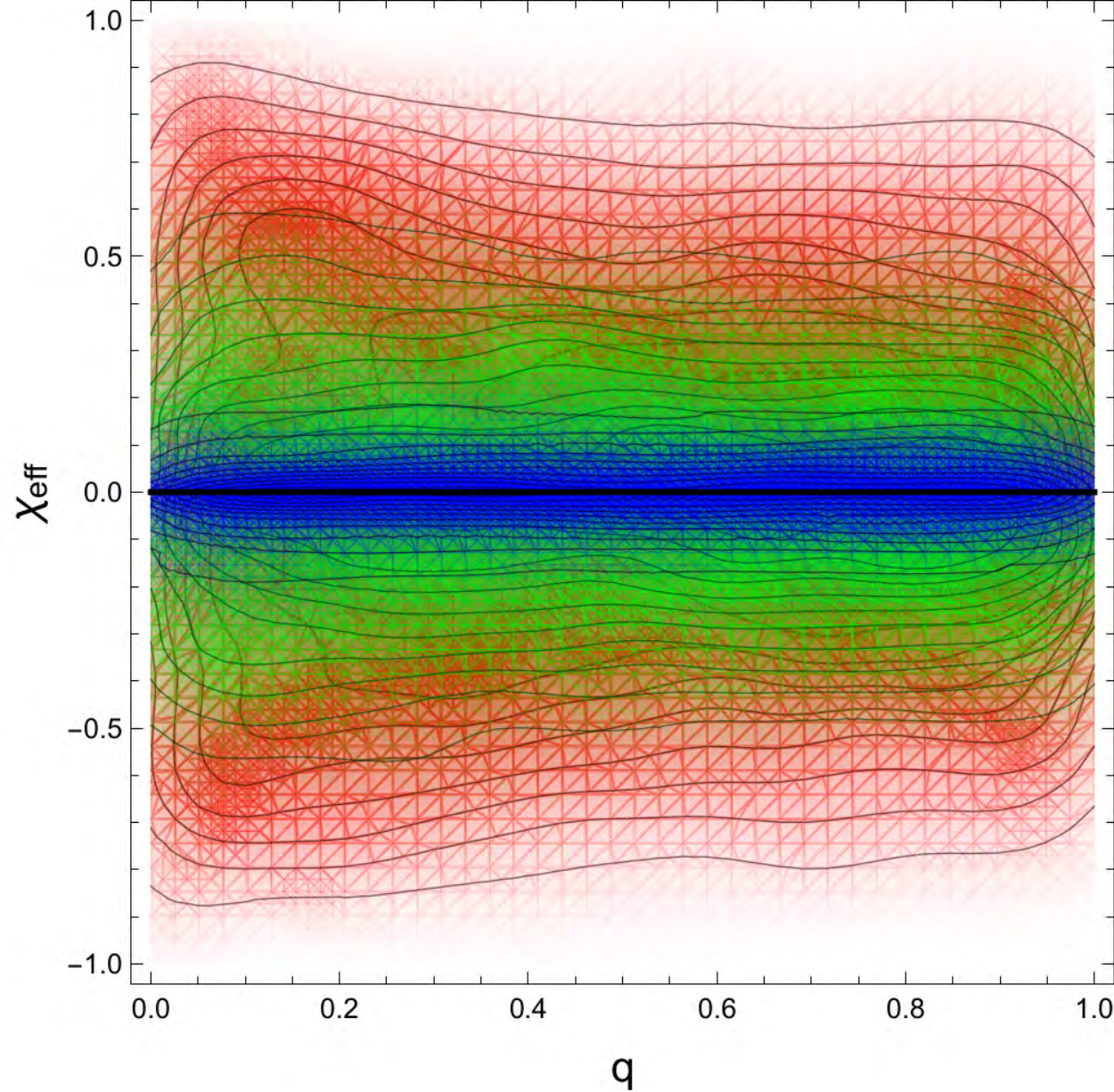}\hspace{1mm}
\includegraphics[width=0.32\textwidth,angle=0]{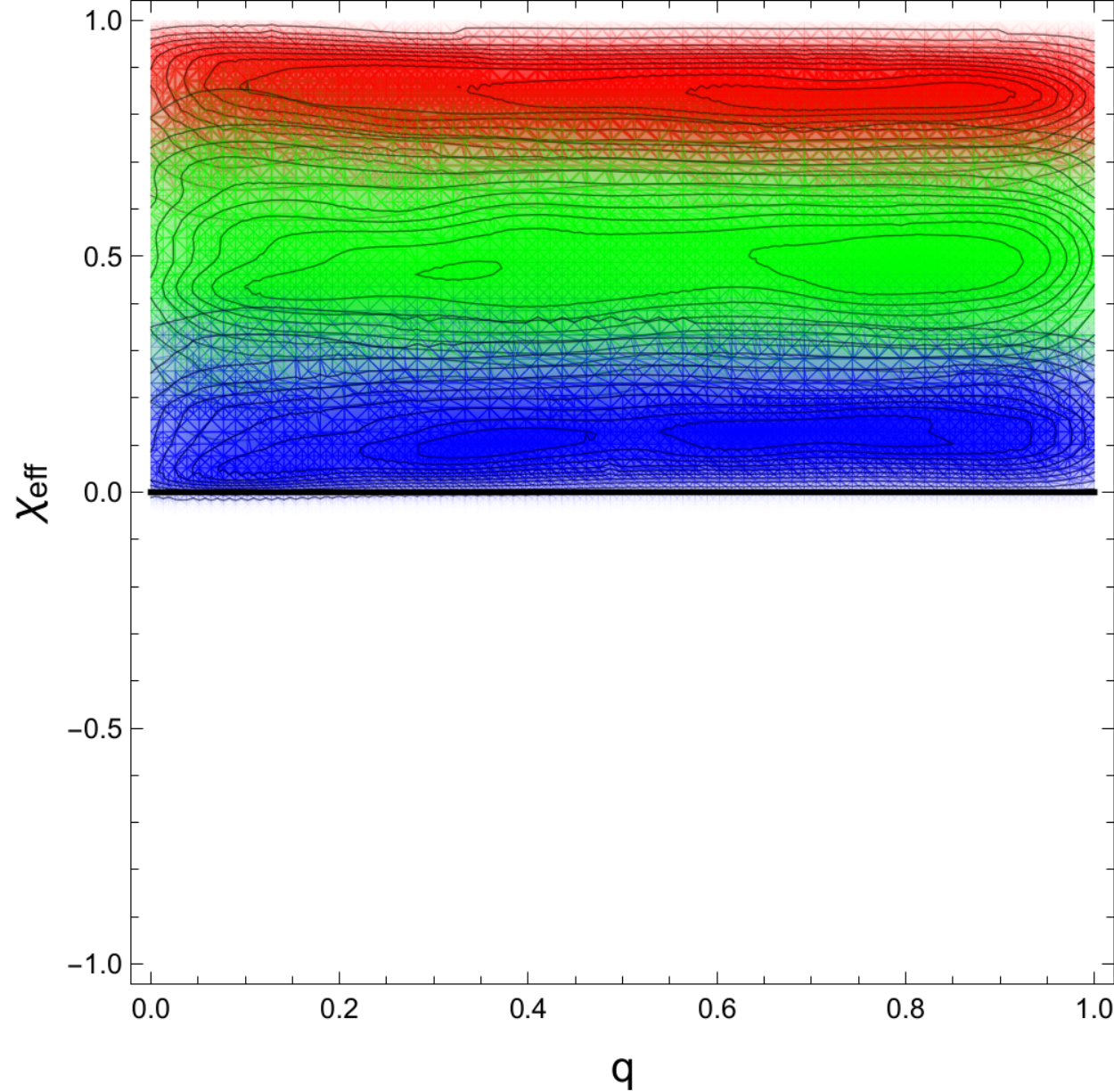}\hspace{1mm}
\includegraphics[width=0.32\textwidth,angle=0]{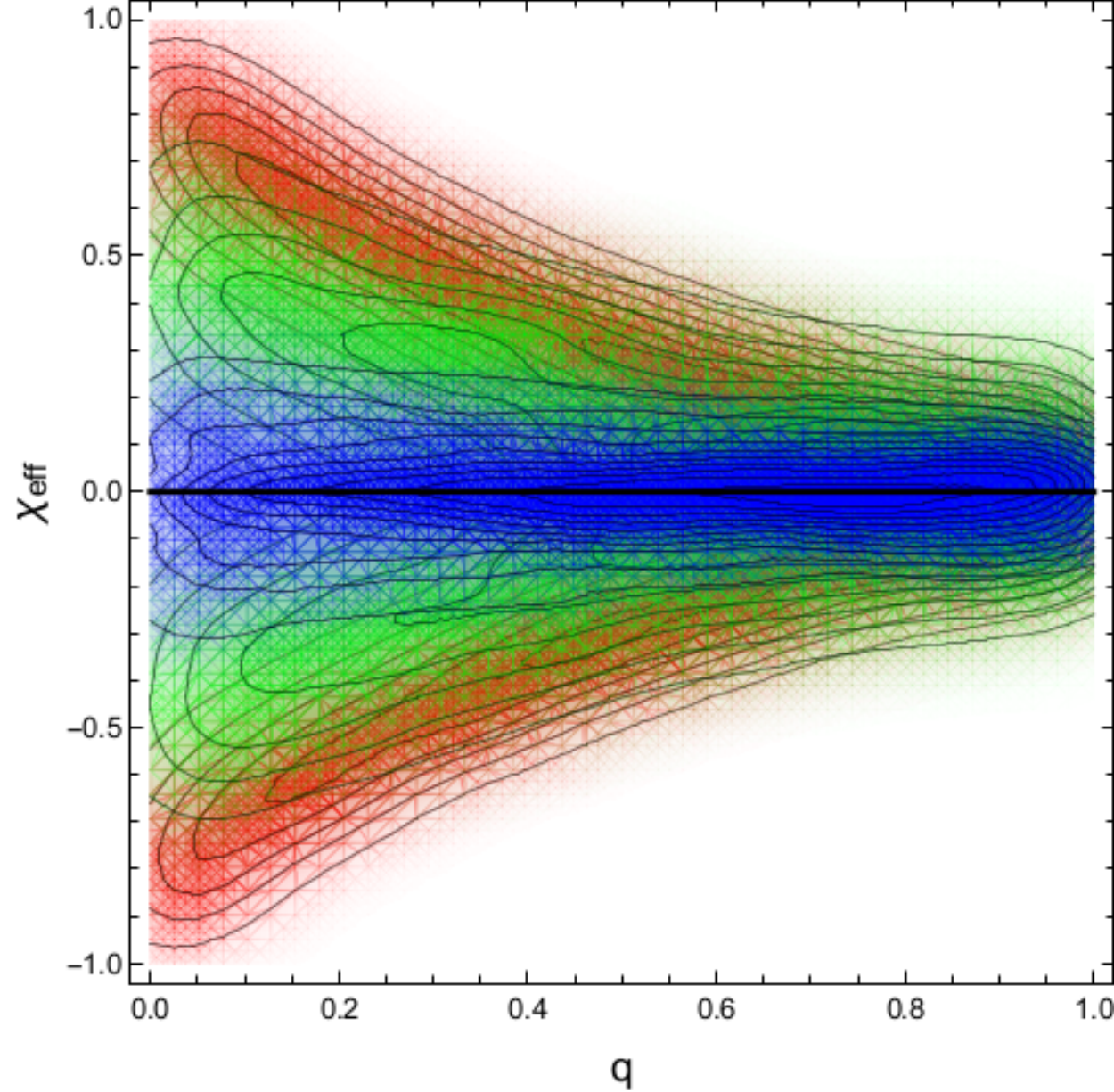}\\[2mm]
\includegraphics[width=0.32\textwidth,angle=0]{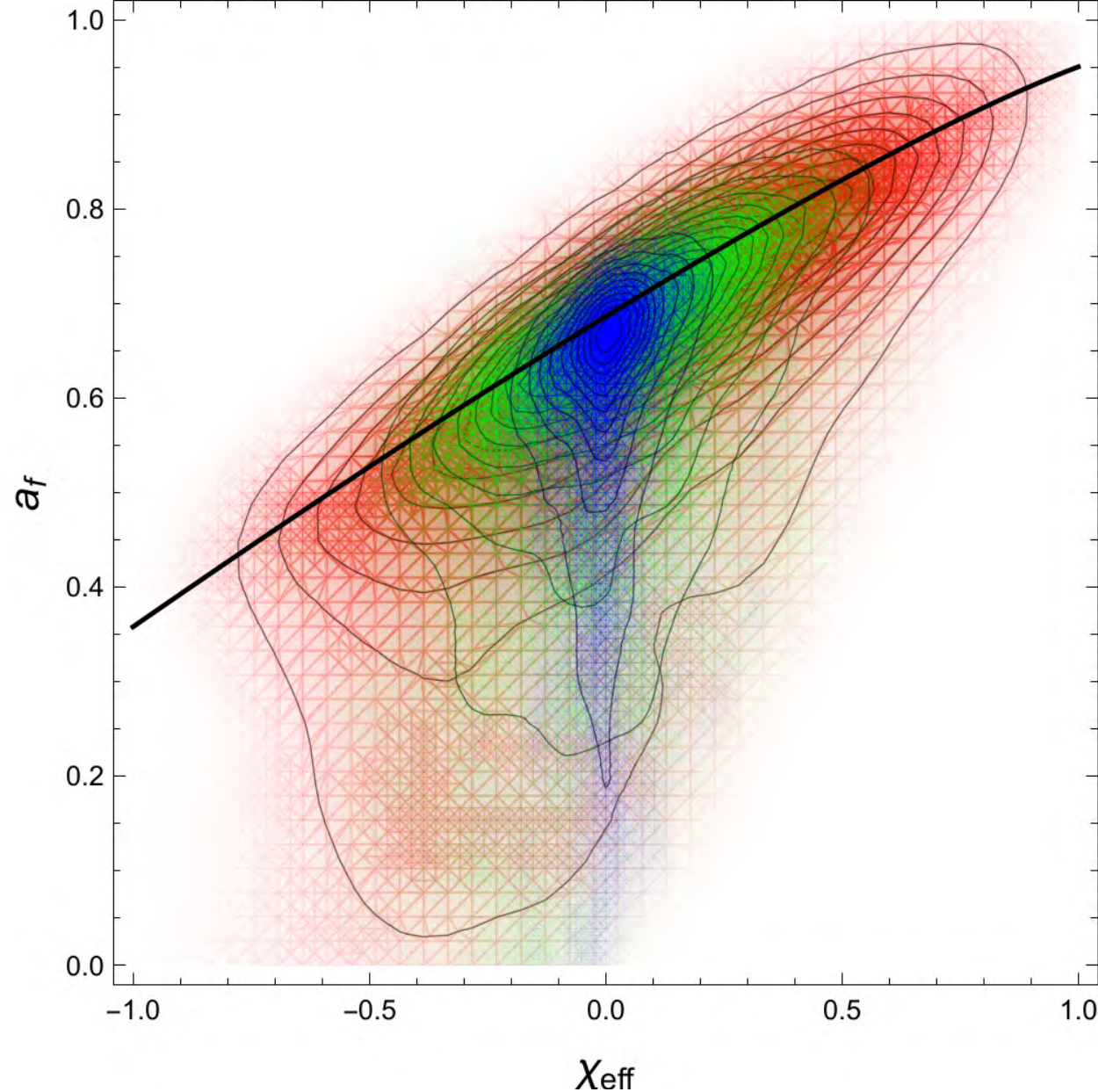}\hspace{1mm}
\includegraphics[width=0.32\textwidth,angle=0]{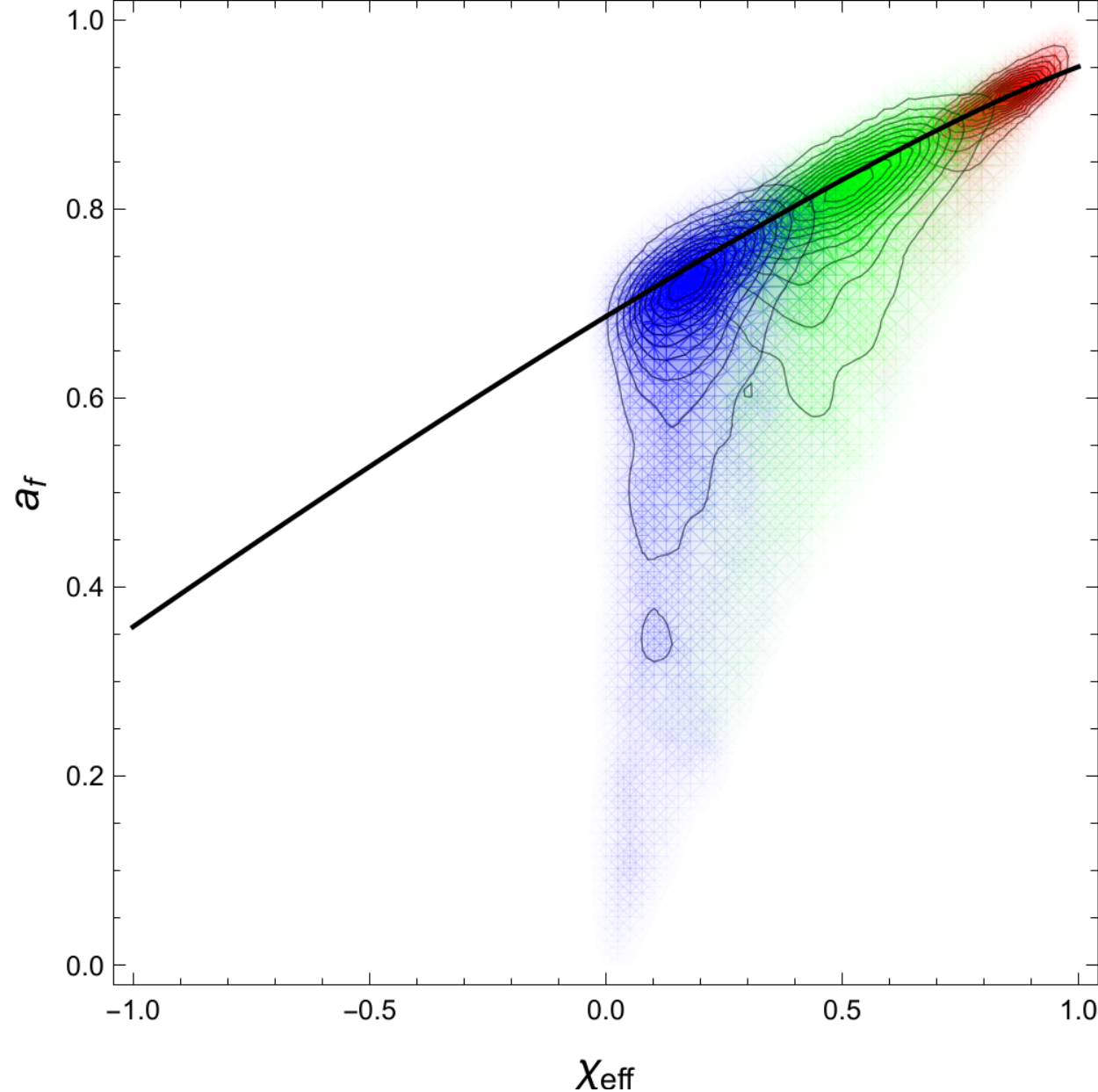}\hspace{1mm}
\includegraphics[width=0.32\textwidth,angle=0]{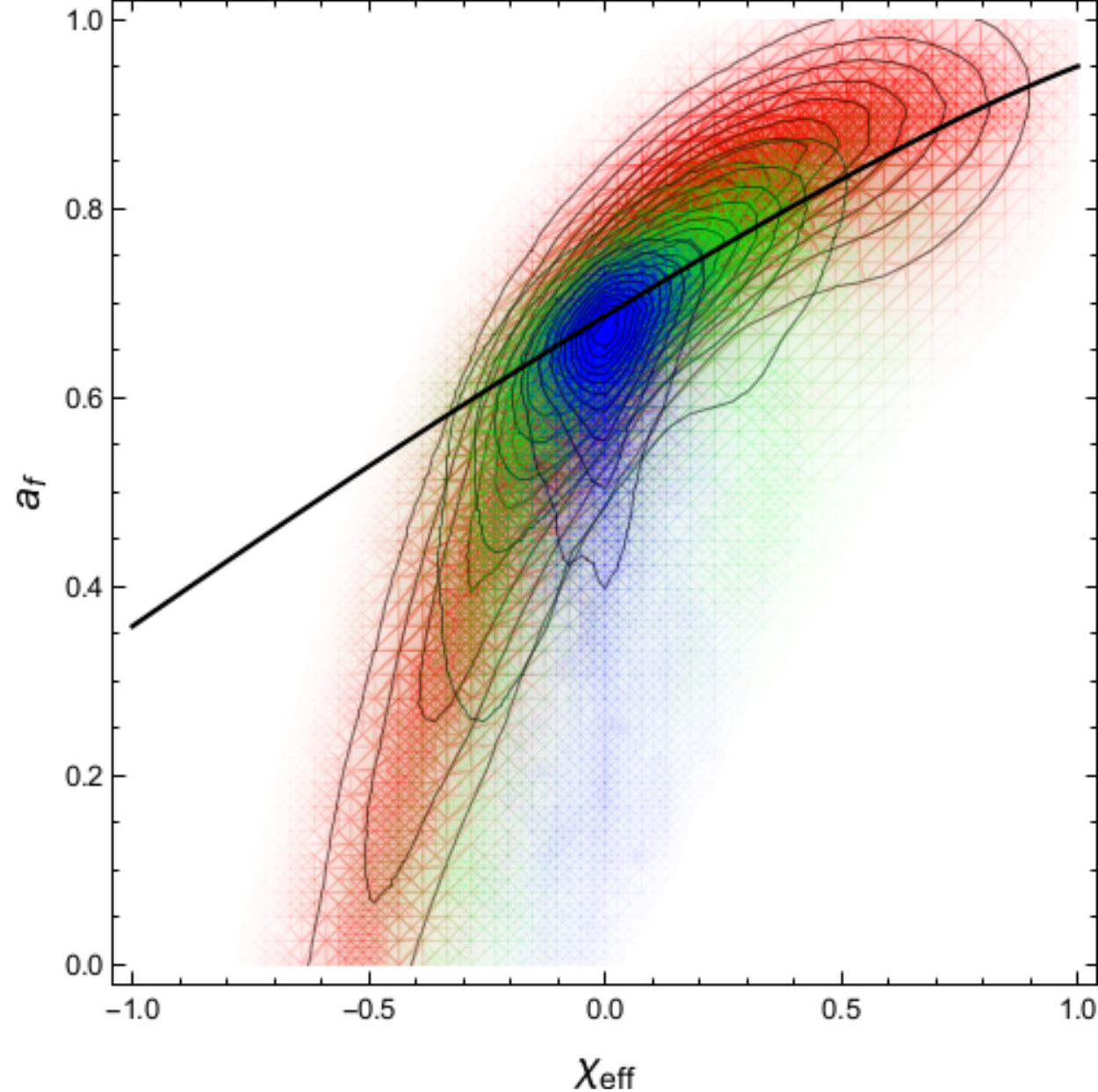}
\caption{The priors in the planes $(q,\,\af)$  (top), $(q,\,\chieff)$ (middle) 
and $(\chieff,\,\af)$ (bottom) for the three hypothesis $H_{i=2,3,4}$: isotropic spin (left), aligned spin (middle) and anti-aligned spin (right), for spins centered at $\mu=0,\,0.5$ and 1 (blue, green and red resp.). 
}
\label{fig:Priors}
\end{figure*}

\section{Bayesian Population  Analysis}

The Bayes theorem relates the likelihood $\L(d_j|\theta)$ of the data $d_j$ for a given set of parameters $\theta$, with the posterior probability of the parameters given the data, $\P(\theta|d_j)$, via the prior knowledge about the parameters of the population model $i$, $\Pi_i(\theta)$,
\bea
\hspace{-2mm}\P_i(\theta|d_j) &=& \frac{\L(d_j|\theta)\,\Pi_i(\theta)}{E(d_j)}\,, \nn \\
E_{ij}(d) &=& \int d\theta\,\L(d_j|\theta)\,\Pi_i(\theta)\,,  \label{Evidence}
\eea
where $E_{ij}(d)$ is the Bayesian evidence for the data $d_j$ and the population model $i$. Here we will consider different priors, $\Pi_i(\theta)$, for the distribution of the parameters in each of the spin population models characterized by the hypothesis $H_i$. 
We first compute the multidimensional likelihood for each BBH event from the LIGO/Virgo published samples, and then calculate the priors associated with each population hypothesis to derive the evidence (\ref{Evidence}).

\subsection{Likelihoods for LIGO/Virgo BBH events}

We calculate the multidimensional likelihoods (i.e. including correlations) from the posteriors and priors provided by the LIGO public documentation page.\footnote{
{\tt https://dcc.ligo.org/cgi-bin/DocDB/DocumentDatabase}} We use the parameter estimation samples for each event, which are given in terms of the fundamental parameters $(m_1,m_2,s_1,s_2, \cos\theta_1, \cos\theta_2)$, and using the expressions of $\chieff$ in (\ref{eq:chieff}) and $\af$ in (\ref{eq:af}) for each sample point, we construct the full multidimensional posterior and prior distributions for the derived parameters $\theta=\{q,\,\chieff,\,\af\}$, marginalizing over the other parameters. Finally we generate the multidimensional likelihoods $\L_{\rm LVC}(d_j|\theta)$ dividing the posteriors $\P_{\rm LVC}(d_j|\theta)$ by their corresponding priors $\Pi_{\rm LVC}(\theta)$. We have plotted the LVC likelihoods for the three parameters $\theta=\{q,\,\chieff,\,\af\}$ in Fig.~\ref{fig:Likelihood}.

When using these likelihoods to infer population properties in our Bayesian analysis, we will not include possible selection effects in the ($\chieff,\,\af,\,q$) variables, like the observational bias towards positive $\chieff$ values described in ~\cite{Ng:2018neg}, or possible effects from the fact that the waveform bank only considers aligned spins. We expect these effects to be small given the spin variables used, but should eventually be considered in a future analysis.

\subsection{Priors on the BBH population}

Once we have chosen to use the subset of parameters $\theta=\{q,\,\chieff,\,\af\}$ for the Bayesian inference, we need to specify the different hypotheses on the spin distribution of the merging BBH population. We have chosen 5 different hypotheses $H_i$ on the spin magnitude and orientation, which in turn depend on ``hyperparameters" $\Lambda = \{\sigma,\,\mu\}$ that determine the width and the mean of the distribution of the spin magnitude.

These five basic Hypotheses are:

\begin{itemize}

\item $H_0$: {\em isotropic spin orientation, \!$\cos\theta_{LS_i}\!\in\![-1,1]$, flat prior on spin magnitude in the range $S_i\in[0,1]$, flat prior on mass ratio, $q\in[0,1]$. This is the "null" hypothesis.}

\item $H_1$: {\em isotropic spin orientation, \!$\cos\theta_{LS_i}\!\in\![-1,1]$, Gaussian prior on spin magnitude with $\mu=0$ and $\sigma\in[0,1]$, flat prior on mass ratio, $q\in[0,1]$.}

\item $H_2$: {\em isotropic spin orientation, \!$\cos\theta_{LS_i}\!\in\![-1,1]$, Gaussian prior on spin magnitude with $\sigma=0.2$ and $\mu\in[0,1]$, flat prior on mass ratio, $q\in[0,1]$.}

\item $H_3$: {\em aligned spin orientation, $\cos\theta_{LS_i}$ sampled from Gaussian centered at $+1$ and width $0.05$, Gaussian prior on spin magnitude for $\sigma=0.2$ and $\mu\in[0,1]$, flat prior on mass ratio, $q\in[0,1]$.}

\item $H_4$: {\em anti-aligned spin orientation, $\cos\theta_{LS_i}$ sampled from Gaussian centered at $\pm1$ and width $0.05$, Gaussian prior on spin magnitude for $\sigma=0.2$ and $\mu\in[0,1]$, flat prior on mass ratio, $q\in[0,1]$.}
 
\end{itemize}

We generate the multidimensional prior probability distributions for each hypothesis $\Pi_i(\theta)$ from $10^5$ random realizations in ($q, \cos\theta_i, \mu, \sigma$), giving rise to the corresponding prior distributions in the $\theta= \{q,\,\chieff,\,\af\}$ parameter space.
To compute these distributions we have used the semianalytic expression of $a_f$ given in equation (\ref{eq:af}) in terms of the fundamental parameters $(m_1,m_2,s_1,s_2, \cos\theta_1, \cos\theta_2)$. 

Fig.~\ref{fig:Priors} shows the projections on the planes $(q,\,\af)$, $(q,\,\chieff)$ and $(\chieff,\,\af)$ of the prior probability distributions $\Pi_i(q,\,\chieff,\,\af)$ for the last three hypotheses $H_{i=2}$ (isotropic spins, left column), $H_{i=3}$ (aligned spins, central column) and $H_{i=4}$  (anti-aligned spins, right column). In each of these plots, we have chosen for illustrative purposes three different choices for the spin magnitude: Gaussian distributions centered at $\mu=0,\,0.5,\,1$ (blue, green and red respectively) with a common width $\sigma=0.2$. The black curves in the $(q,\,\af)$ and  $(q,\,\chieff)$ planes corresponds to the limit case of zero spin of the underlying black hole population, while the black curve in the $(\chieff,\,\af)$ plane corresponds to the limiting case of $q=1$.

These plots show that the different hypotheses on the underlying spin configurations populate very different areas of the $(q,\,\chieff,\,\af)$ parameter space, and therefore, taken as priors and integrated over the whole parameter space with the likelihood of each event shown in Fig.\ref{fig:Likelihood}, we expect very different Bayesian evidence and therefore significantly informative Bayes ratios for the different hypotheses. 

Comparing the distributions on the different planes we see that while the $\chieff$ distributions are flat with respect to $q$, a well known degeneracy for random spins, the $\af$ distributions have a strong dependence on $q$. Models with high and aligned or anti-aligned spins simply cannot produce events with low $\af$ and $q$ values, as found for event GW190814. Only populations with close to zero spin or high but isotropic spins can populate the lower left corner of the $(q,\af)$ plane. If the forthcoming O3a catalog includes more events of low $q$, we expect a much better determination of the spin of the underlying black hole population.

These five hypotheses give rise to multidimensional priors in the space of derived parameters 
$\theta= \{q,\,\chieff,\,\af\}$, which can then be used to integrate them, together with the LVC likelihoods, to obtain the Bayesian evidence for each prior hypothesis $H_i$, which depend on $\Lambda=\{\sigma,\,\mu\},$
\be\label{eq:EviParam}
E_{ij}(\Lambda) = \!\!\int\!d^3\theta\,
\L(h_j|\theta)\,\Pi_i(\theta|\Lambda)\,.
\ee

Once we have obtained the LVC likelihoods from their parameter estimation samples, and computed the individual priors for our hypothesis, we can perform the 3D integration with Mathematica.

\begin{figure*}
\centering 
\includegraphics[width=0.5\textwidth,angle=0]{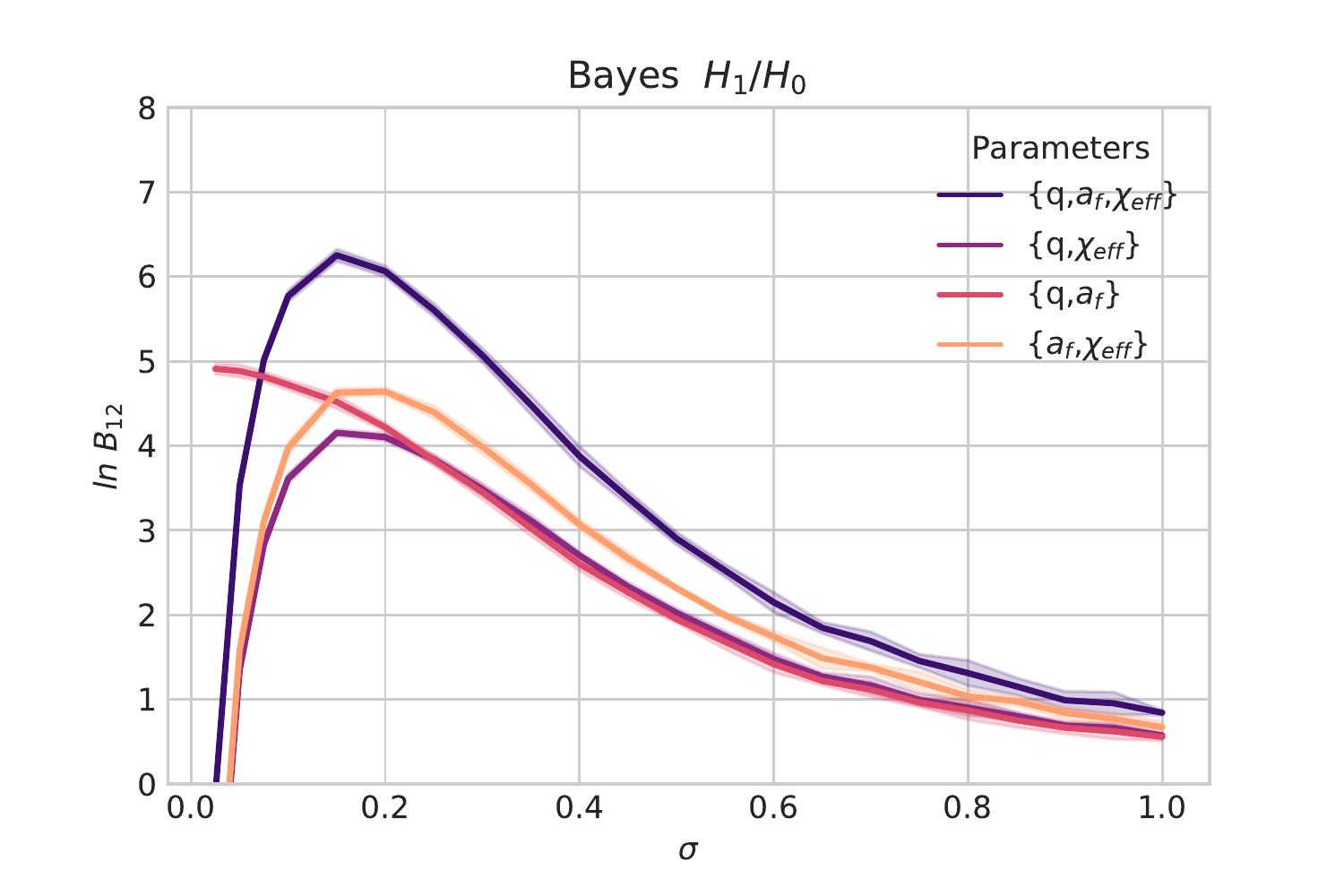}
\includegraphics[width=0.5\textwidth,angle=0]{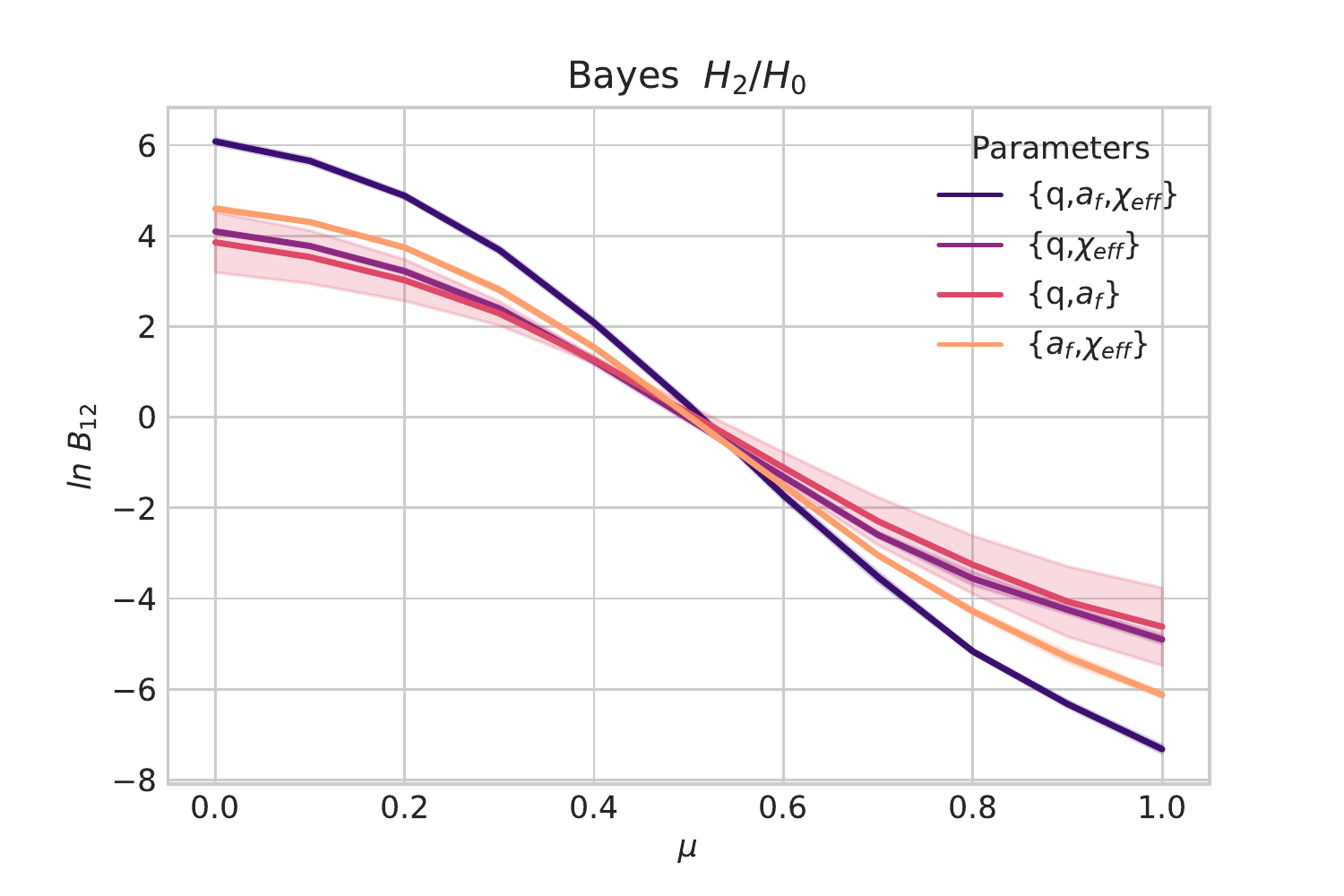}\\[3mm]
\includegraphics[width=0.5\textwidth,angle=0]{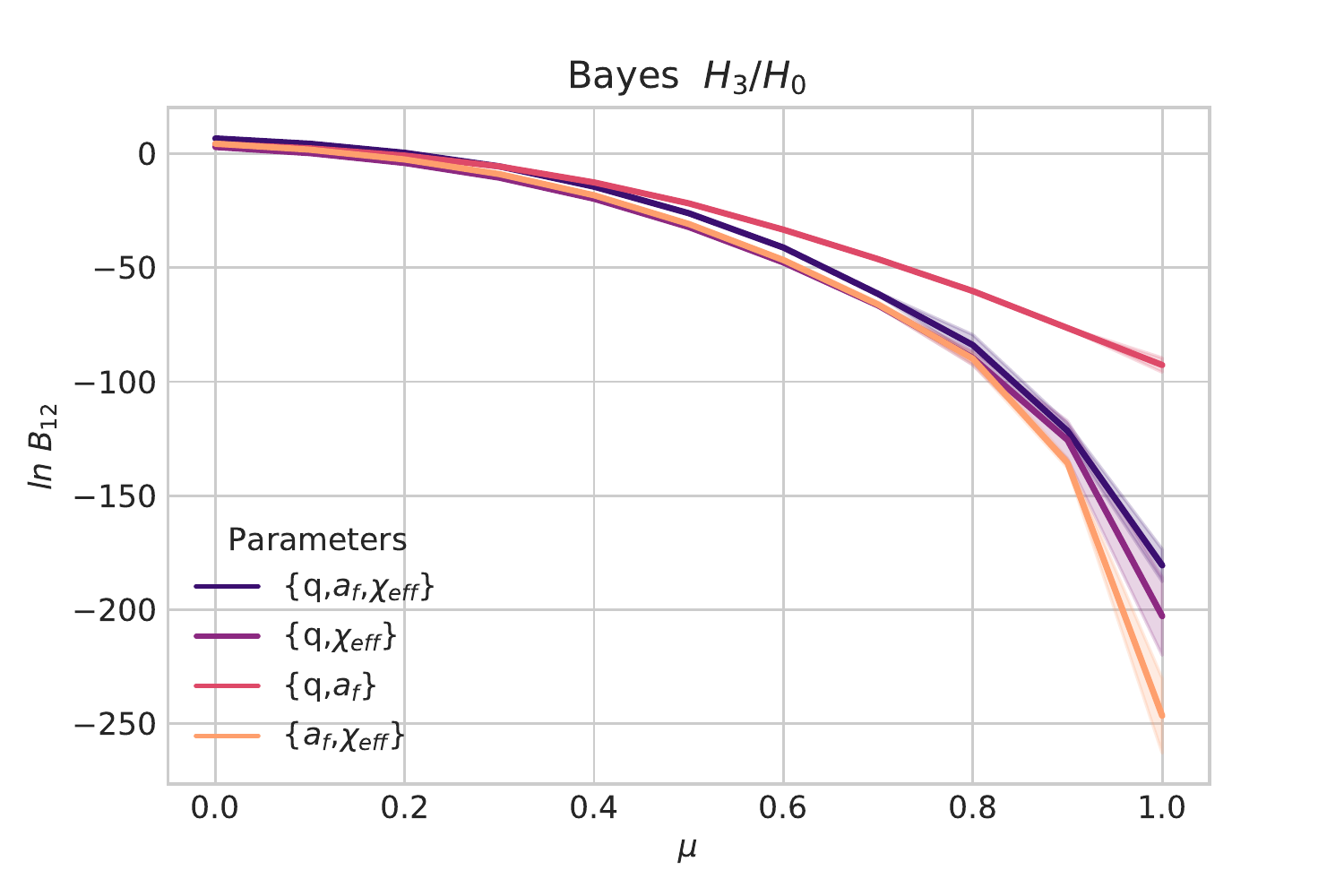}
\includegraphics[width=0.5\textwidth,angle=0]{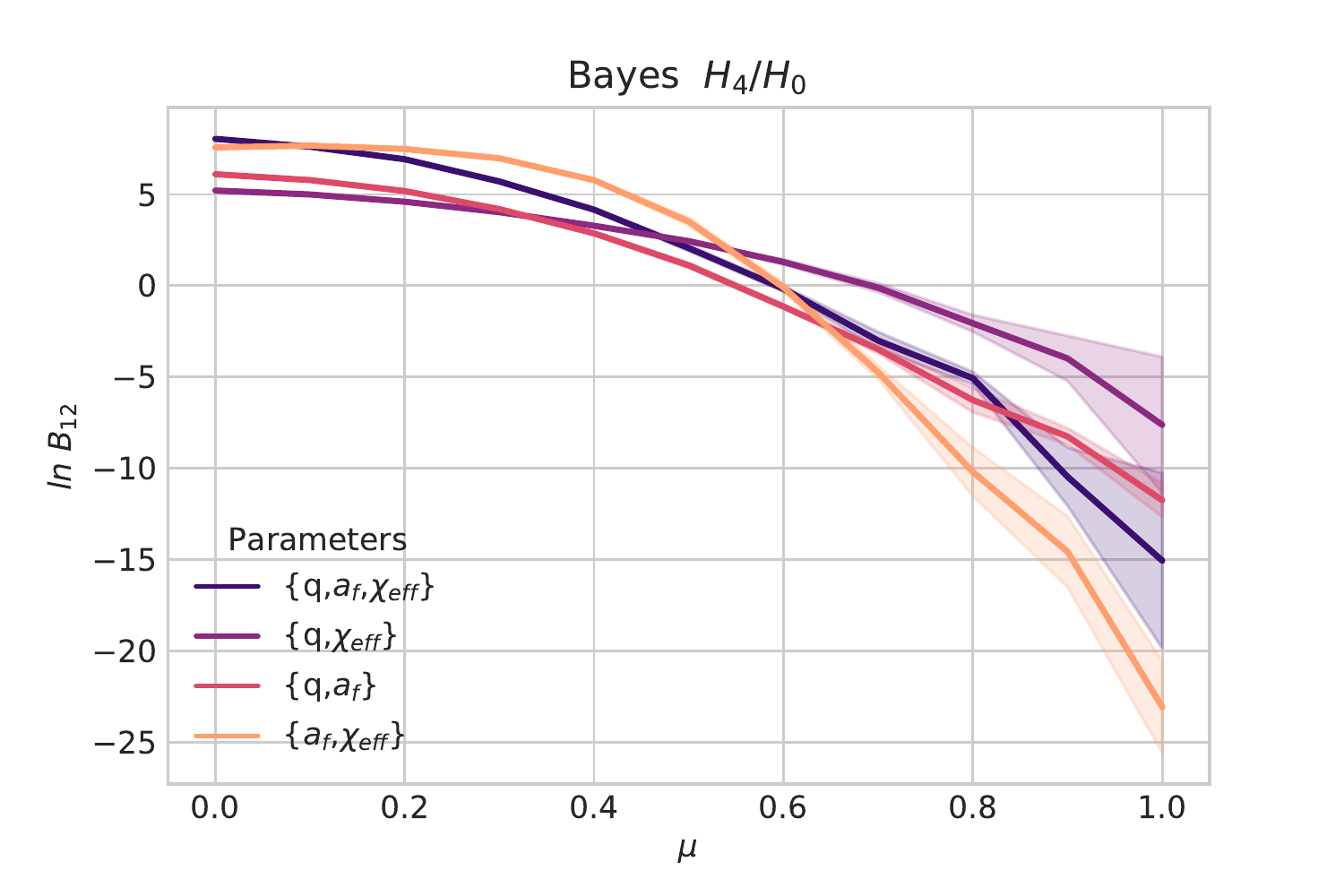}
\caption{The log of the Bayes ratios in the variables $(q,\,\chieff,\,\af)$ for the four hypothesis: $H_1$ = isotropic spin centered at zero and width $\sigma\in[0,\,1]$ (top left) and $H_{2,3,4}$ = isotropic spin (top right), aligned spin (bottom left) and anti-aligned spin (bottom right), for spin distributions centered at $\mu\in[0,\,1]$, for $\sigma=0.2$. We also study the sensitivity of the Bayes ratios to the evidence in the 2D planes $(q,\,\chieff)$, $(q,\,\af)$ and $(\chieff,\,\af)$, with increasing values of the Bayes ratios.
We find strong evidence  $|\ln B_{12}| \geq 6$ at small values of the spin, for all four Hypotheses $H_i$ versus the (null) ``All Flat" Hypothesis $H_0$, attaining the maximum value of $|\ln B_{12}| \geq 8$ for $H_4$ at $\mu \leq 0.01$.
The events contributing to these Bayes factors are the GWTC-1 (10-event catalog) plus the four published run-O3 events from LIGO/Virgo Collaboration. 
}
\label{fig:BayesRatios}
\end{figure*}

\section{Bayes Factors}

In order to evaluate the global significance of a prior hypothesis from the full BBH catalog, we compute the global Bayes factor for the whole BH population. For this, we will multiply the individual Bayes ratios, assuming that all events in the catalog are independent,
\be\label{eq:BayesRatio}
\ln\,B^i_{12}(\Lambda) = \sum_{j=1}^N \ln\frac{E_{ij}}{E_{0}}(\Lambda)\,,
\ee 
for each hypothesis $H_i(\Lambda=\{\sigma,\,\mu\})$.

In practice what we do is a reweighting of the priors, using our population model hypothesis, $\Pi_i(\theta)$, versus the published LIGO/Virgo priors, $\Pi_{\rm LVC}(\theta)$, which depend on the assumptions of the experiment on each event,
\bea
E_{ij}(\Lambda) &=& \int d^3\theta\,\L_{\rm LVC}(d_j|\theta)\,\Pi_i(\theta|\Lambda) \nn\\ \label{eq:EvidenceLVC}
&=& E_{\rm LVC}\,\int d^3\theta\,\P_{\rm LVC}(\theta|d_j)\,\frac{\Pi_i(\theta|\Lambda)}{\Pi_{\rm LVC}(\theta)}\,.
\eea
Since $E_{\rm LVC}$ is the same for all hypotheses, it factors out in the Bayes ratio ($\ref{eq:BayesRatio}$). We can then perform the integral (\ref{eq:EvidenceLVC}) in the full 3D parameter space $\theta=\{q,\,\chieff,\,\af\}$ maintaining all the correlations in the multivariate priors and posteriors.

We show in Fig.~\ref{fig:BayesRatios} the Bayes ratios of the four different hypotheses $H_{i=1,2,3,4}$ with respect to the null hypothesis $H_0$ of flat distributions for mass ratios, spin magnitudes and random spin orientations. 
According to Jeffrey's scale~\cite{Jeffreys:1939xee}, when $\ln\,B_{12} \geq 5$, hypothesis 1 is significantly more likely that hypothesis 2, see however~\cite{Nesseris:2012cq}. 

It is clear from figure~\ref{fig:BayesRatios} that low spin magnitudes are significantly preferred for the BBH population of LIGO/Virgo events, with Bayes ratios above $\ln B_{12}=5$ for $\sigma,\mu<0.2$, therefore we find strong evidence for small values of the spin, for all four Hypotheses $H_i$ versus the "null" (All Flat) Hypothesis $H_0$. In the case of Hypothesis $H_1$, whith zero spin and allowing for variable width, we find strong evidence ($\ln B_{12} = 6.2$) for relatively narrow spin distributions $\sigma=0.15$. This low and narrow spin hypothesis represents the isotropic spin distribution that one would expect from an underlying population of primordial black holes. On the other hand, when we vary the spin magnitude as a Gaussian centered at $\mu$ for fixed width $\sigma=0.2$, we find, in all three orientation Hypothesis $H_{2,3,4}$, that the maximum evidence occurs again for  $\mu = 0$. Therefore we conclude that, whatever the orientation, there is very strong evidence for low spins in LIGO/Virgo BH. 

Note that the aligned spin hypothesis $H_3$ has Bayes factors $\ln B_{12} < -5$ for spins $\mu>0.25$, reaching large Bayes ratios $\ln B_{12} < -100$ for large spins $\mu>0.8$, therefore our analysis of LVC catalog strongly {\em disfavoures} a population of BBH with aligned spins and magnitudes greater than $0.25$. Note also that hypothesis $H_1$ and $H_4$, for isotropic and anti-aligned spins, have a similar behaviour in their Bayes factors at low spins, $\mu<0.5$, reaching values above $\ln B_{12} \simeq 6$ for $\mu<0.06$ in the isotropic case, and above $\ln B_{12} \simeq 8$ for $\mu<0.02$ in the anti-aligned case, giving slightly higher significance for anti-aligned versus isotropic spins in the range $\mu<0.5$. However, for large spins, $\mu>0.5$, the anti-aligned hypothesis is much more strongly disfavoured than the isotropic one.

We also compute the Bayes factors for the three pairs ($q,\,\chieff$), ($q,\,\af$) and ($\chieff,\,\af$) finding less evidence for low spins compared with the full analysis. The least significant is the Bayes ratio in the plane ($q,\,\chieff$), giving only mild preference for low spins. As we include information coming from $\af$, as in ($\chieff,\,\af$), ($q,\,\af$) and ($q,\,\chieff,\,\af$), the evidence for low spin rises, in some cases up to $\ln B_{12}\sim 6$ or above, which is very strong evidence in favor of that hypothesis.

Historically, the emphasis has been focused on the importance of distinguishing between aligned and anti-aligned astrophysical models of BBH formation. Now that the evidence for small spins is so strong, the orientation of the spins becomes less relevant. It is much more interesting to characterize the posterior distributions of the whole population and to quantify the deviation from zero spins according to the various hypotheses $H_i$, which take into account orientation. This will be useful in order to characterize the possible origin of LVC BBH events from primordial black hole populations.


\begin{figure*}
\centering 
\includegraphics[width=0.32\textwidth,angle=0]{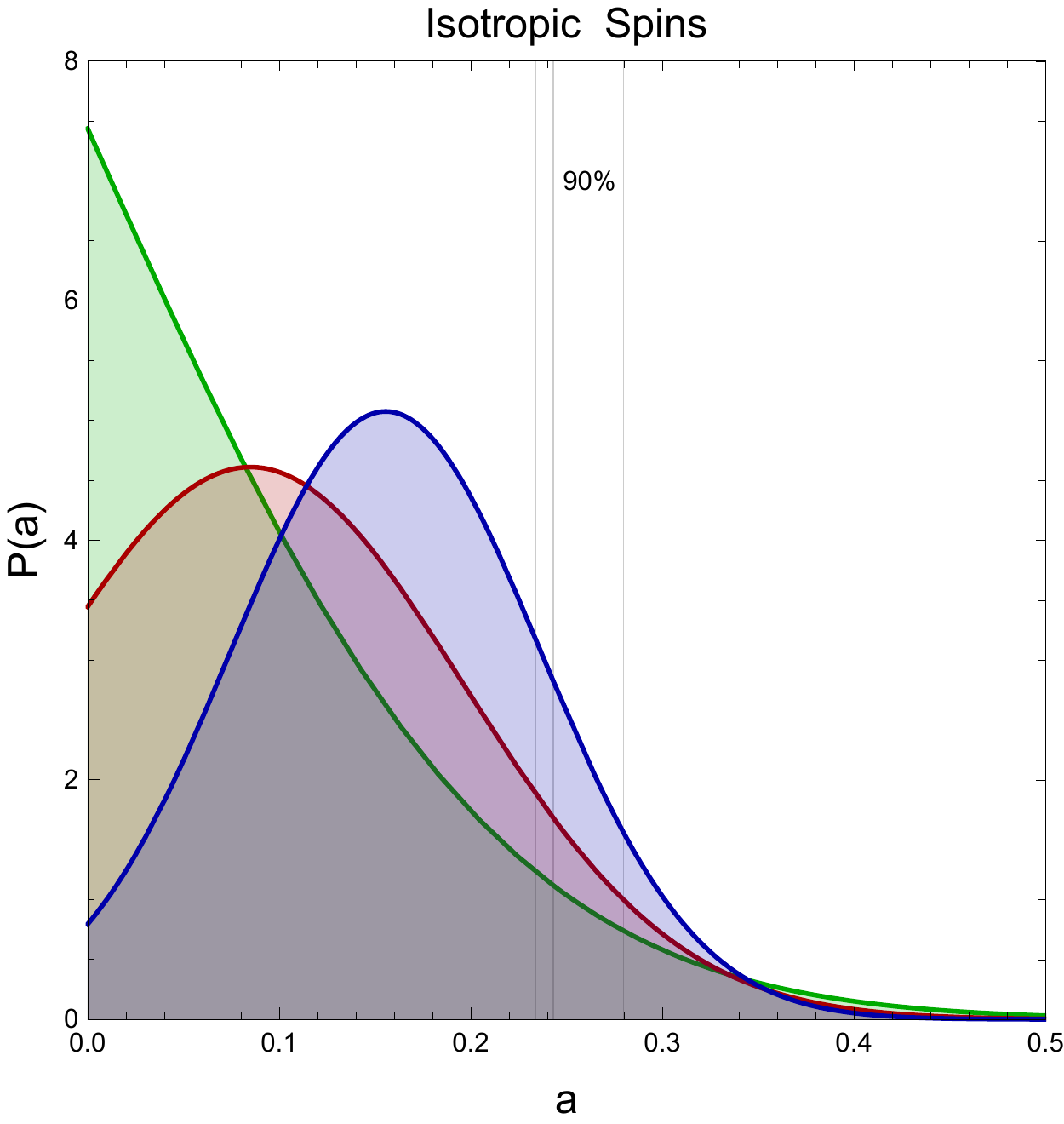}
\includegraphics[width=0.32\textwidth,angle=0]{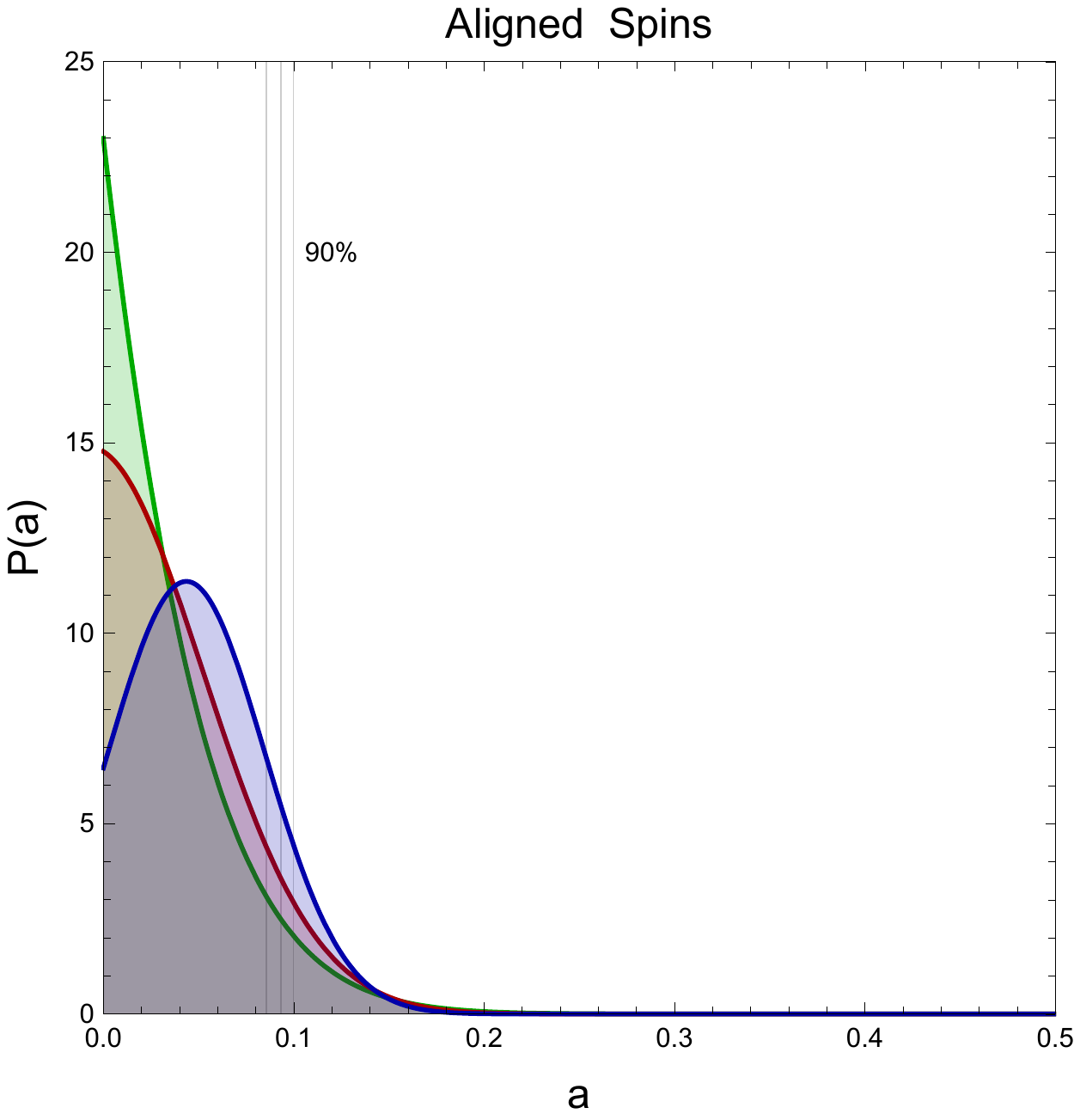}
\includegraphics[width=0.32\textwidth,angle=0]{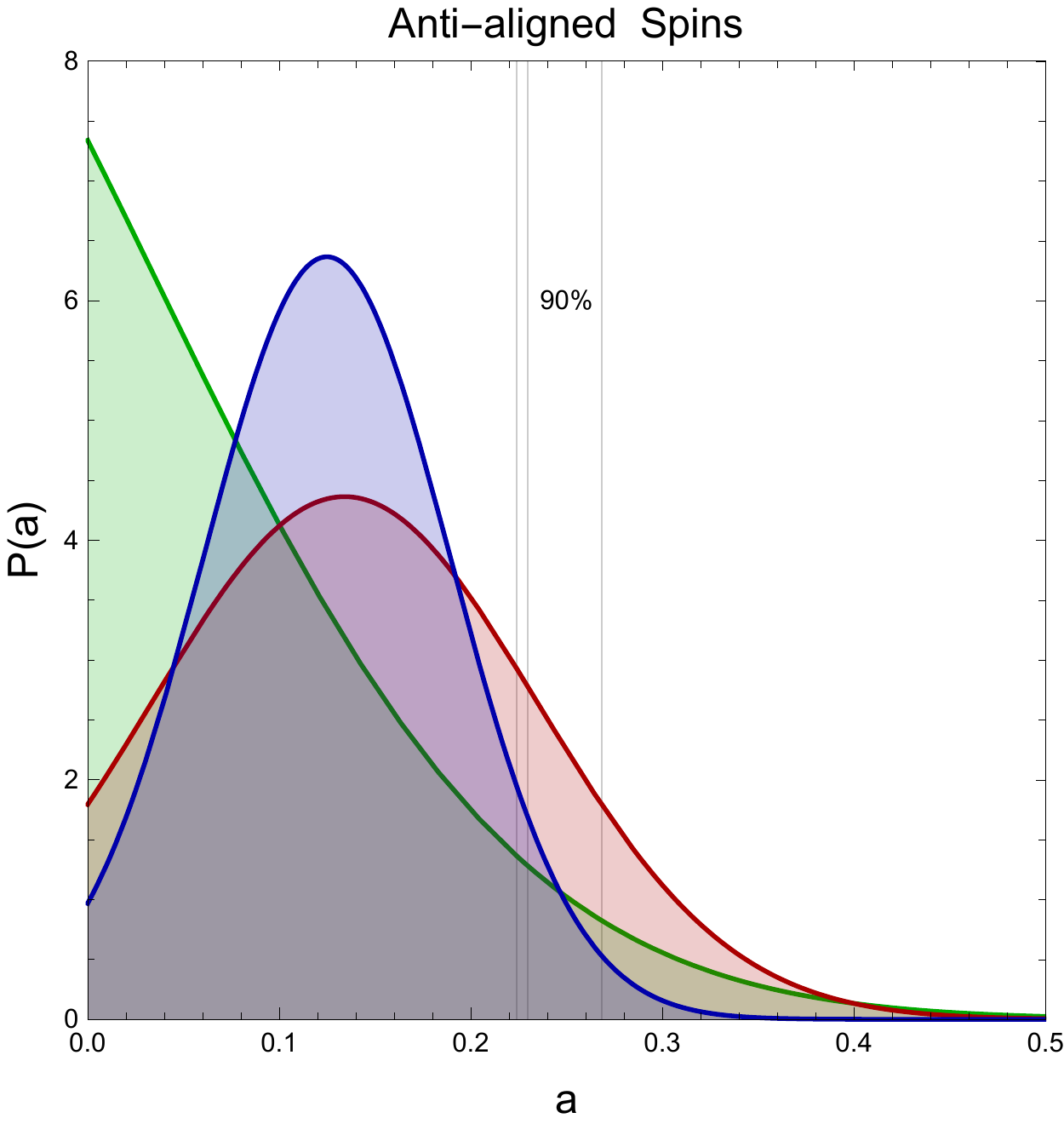}
\caption{The posterior PDFs $P(a)$ for the spin magnitude in the case of isotropic (left), aligned (center) and anti-aligned (right) hypothesis. We have computed the PDFs for three different spin widths, $\sigma=0.05$ (blue), $\sigma=0.1$ (red) and $\sigma=0.25$ (green). The PDFs do not seem to depend much on the width $\sigma$ of the distribution. In all cases, the PDFs peak close to zero spin, specially for $\sigma=0.25$. The vertical gray lines correspond to the 90\% c.l. limit, which is almost the same for all PDFs.
}
\label{fig:PDF-hyp}
\end{figure*}

\section{Bayesian hierarchical modeling}

Given that the random isotropic spin hypothesis, $H_1$, gives a value of $\ln\,B_{12} \sim 6$, strongly favouring low spin magnitudes $a<0.2$ for the combined LVC BBH events, we would like to explore in detail the black hole population distribution of spin magnitudes.

In order to go forward in the information content with respect to the Bayesian approach of the previous section, we will use Bayesian hierarchical modeling, see e.g.~\cite{Talbot:2017yur}, as a way to estimate the posterior distribution for spin magnitudes of the LVC black hole population. Here the spin magnitude $a$ enters indirectly through the dependence of other measured quantities, like the effective spin $\chieff$ or the final spin $\af$, on it. Then the posterior PDF for $a$ is computed as 
\be
P_i(a) = \Pi(a)\,\prod_{j=1}^N E_{ij}(a)\,,
\ee
where $E_{ij}(a)$ are the Bayesian evidences (\ref{eq:EviParam}) computed from the LVC likelihoods with the spin-dependent priors $\Pi_i(q,\,\chieff,\,\af|\Lambda)$, with $\Lambda=\{\sigma=0.05,0.1,0.25,\;\mu=a\}$, and $\Pi(a)$ is assumed here to be flat in spin magnitude $a$, not to give any prior preference for any spin magnitude. We have plotted in Fig.~\ref{fig:PDF-hyp} the posterior distributions for the three spin Hypothesis (isotropic, aligned and anti-aligned). In all cases, the preferred spin magnitude is below $a<0.12$, within 50\% c.l. For example, for isotropic spins, currently favoured by the data, we find $a<0.25$ at 90\% c.l.
For aligned spins the PDF is even more strongly peaked around zero spin, with $a<0.1$ at 90\% c.l. All this suggests that the inclusion of the four published run O3 events, with increased sensitivity of the detectors, has revealed a property of the population of LIGO/Virgo black holes that was not present in previous studies~\cite{LIGOScientific:2018jsj}.

\section{Conclusions}

The routine detection of BBH inspirals by LVC has opened the door to a detailed exploration of the nature of black holes and their populations. The online availability of the strain time-streams and the parameter estimation samples for each event allows for an independent analysis, opening the possibility to compute multivariate likelihoods for combinations of derived parameters like the mass ratio, or the effective spin.

It is then a matter of personal choice which parameters to use in order to infer properties of the populations of black holes detected by LVC. In this paper we have concentrated on just three parameters ($q,\,\chieff,\,\af$) that we believe capture the essence of the spin nature of the population of LVC black holes.

We have put forward four different spin-magnitudes and spin-orientation prior hypothesis, $H_{i=1,...,4}$ (i.e. isotropic, aligned and anti-aligned), to compare with the (null) all-flat prior hypothesis, $H_0$, and conducted a Bayesian analysis study to determine the goodness of a given spin-distribution hypothesis for the whole population of LIGO/Virgo black holes.

We find that all spin-orientation hypothesis have a larger Bayes factor for low spins. In some cases the log of the Bayes factor reaches values well above five, thus signalling a strong evidence in favour of low spins (below magnitude $a=0.2$). Moreover, the largest Bayes factors are obtained by spin distributions peaked at zero spin (Bayes factors above 5 for widths $\mu < 0.2$), with very small width.

We also note that aligned spins are strongly disfavoured, specially for large spin magnitudes (with the log of the Bayes factors as low as $-100$), as would be expected from astrophysical black holes from isolated binaries. On the other hand, we find that LIGO/Virgo black hole population has a preference towards low spins with isotropic orientations, consistent with what one would expect from primordial black holes in clusters~\cite{Clesse:2016vqa}.

We have then computed the posterior PDF for the spin {\em magnitude} in the case of the three alternative spin hypothesis, for different spin widths ($\sigma=0.05,\,0.10,\,0.25$). In all cases the spin distribution is peaked at very low values ($a\simeq0.1,\,0.05,\,0.1$ for isotropic, aligned and anti-aligned cases, respectively), which clearly indicates a preference of the whole LIGO/Virgo BH population for low spins, irrespective of orientation.

Therefore, we conclude that only using the BH spin as a discriminator between the astrophysical versus primordial nature of LIGO/Virgo black holes, when considered as a homogeneous population of black holes, our analysis seems to suggest very strongly that they are consistent with being primordial.

\

\section*{Acknowledgements}
JGB and ERM thank the Theory Division for their generous hospitality during their Sabbatical at CERN in 2018, when this work originated. The authors acknowledge support from the Research Project PGC2018-094773-B-C32 (MINECO-FEDER) and the Centro de Excelencia Severo Ochoa Program SEV-2016-0597.

\bibliography{PBH-Spin-Bayes}

\end{document}